\documentstyle{article}

\newcommand{\lbl}[1]{\label{#1}}

\newcommand{\hs}{{}}
\newcommand{\ds}{{}}
\newcommand{\cs}{{}}

\newcommand{\be}{\begin{equation}}
\newcommand{\ee}{\end{equation}}
\newcommand{\bea}{\begin{eqnarray}}
\newcommand{\eea}{\end{eqnarray}}
\newcommand{\half}{{1 \over 2}}

\newcommand{\dd}{{\,{\rm d}}}
\newcommand{\ddx}{{\,{\rm d}^3{\bf x}}}

\newcommand{\imag}{{\rm i}}

\newcommand{\sB}{{\sf B}}

\newcommand{\bk}{{\bf k}}

\newcommand{\bn}{{\bf n}}
\newcommand{\bp}{{\bf p}}
\newcommand{\br}{{\bf r}}

\newcommand{\bu}{{\bf u}}
\newcommand{\bx}{{\bf x}}
\newcommand{\by}{{\bf y}}
\newcommand{\bz}{{\bf z}}
\newcommand{\btu}{{\bf \tilde{u}}}
\newcommand{\kk}{{\rm k}}

\newcommand{\te}{{\tilde{e}}}

\newcommand{\tu}{{\tilde{u}}}

\begin{document}

\title{Modelling of Phase Separation \\
in Alloys with Coherent Elastic
Misfit\thanks{Dedicated to John W. Cahn, on the occasion of his 70th birthday}}
\author{\normalsize Peter Fratzl$^1$, Oliver Penrose$^2$ and Joel
L. Lebowitz$^3$ \\ 
\\ \mbox{} \small
$^1$ Erich Schmid Insitute of Materials Science, Austrian Academy of Sciences,
\\ \small and University of Leoben, Jahnstr. 12, A-8700 Leoben, Austria \\
\small $^2$ Department of Mathematics, Heriot-Watt University, \\
\small Riccarton, Edinburgh EH14 4AS, Scotland, UK \\
\small $^3$Departments of Mathematics and Physics, Rutgers University, \\
\small Hill Center,
Busch Campus, New Brunswick, 08903 New Jersey, USA \\  \mbox{}}

\maketitle

\section*{Abstract}
                           
\cs Elastic interactions arising from a difference of lattice spacing 
between two coherent phases can have a strong influence on the 
phase separation (coarsening) of alloys. 
If the elastic moduli are different in the two phases, the elastic
interactions may accelerate, slow down or even stop the
phase separation process. If the material is elastically anisotropic, 
the precipitates can be shaped like plates or needles instead of 
spheres and can form regular precipitate superlattices. 
Tensions or compressions applied externally to the specimen
may have a strong effect on the shapes and arrangement of the precipitates.
In this paper, we review the main theoretical approaches that
have been used to model these effects and we relate them to experimental 
observations. The theoretical approaches considered are
(i) `macroscopic' models treating the two phases as elastic 
media separated by a sharp interface (ii) `mesoscopic' models 
in which the concentration varies continuously across the
interface (iii) `microscopic' models which use the positions 
of individual atoms.

\bigskip\bigskip
\noindent

Keywords:  kinetics of phase separation, quenched alloys, elastic
interactions, sharp interface model, diffuse interface models, atomic
lattice models

\newpage
\tableofcontents
\newpage

\section{Introduction}

\subsection{The scope of this paper}\lbl{scope}

The separation of an alloy into two phases 
is a technologically important phenomenon
whose theory owes much to the work of John Cahn.
When the two phases first form they are very finely mixed,
with large amount of interface between the two phases.
Atoms then diffuse within the material so as to reduce
the amount of interface, making the mixture of phases
more and more coarse as time proceeds. The driving force
for this coarsening process comes from the free energy
of the interface. For many
alloys, where the sizes of the two kinds of atoms
composing the alloy are not too different, the atoms
manage to fit themselves on to a common lattice,
but they can achieve this only at the cost of
significant distortion of the lattice.
At first, the energy arising from this distortion
is relatively unimportant, but as the coarsening
proceeds the interfacial energy decreases and
eventually the elastic 
\cs \hs interactions, which are of long range and can
also be strongly anisotropic,
may predominate.
This change in the dominant driving force
can dramatically affect both the rate of coarsening and
the structure of the domains themselves
in the late stages of the coarsening process.

In the present article we review some theoretical approaches
that have been used to model the effects of elastic interactions on
coarsening in alloys. 
We shall confine ourselves to alloys which are coherent
(meaning that the lattice structure is only distorted, not
disrupted, by the misfit between the two types of atom),
and for that reason we shall not consider any effects
involving dislocations, such as plasticity. Moreover,
we shall consider only first-order phase transformations, in which
the concentrations are different in the two phases,
so that diffusive transfer of material from one place
to another is an essential part of the process;
martensitic transformations, for example, will not be considered.

An idea of the relative importance of the elastic and surface energy
effects can be obtained by comparing the surface energy of
a spherical precipitate of radius $R$,
which is $4\pi \sigma R^2$  where $\sigma$ is the interfacial
energy per unit area (surface tension),
with the elastic energy due to such a precipitate, which is of order
$\half (4\pi R^3/3) G q^2$ where $G$ is a typical elastic modulus,
e.g. the shear modulus, and
$q$ is a typical strain due to the misfit.
The two energies are equal when
$R \sim R_0 = 6 \sigma / G q^2$. At first, the precipitates are much
smaller than this size, so that the interfacial tension predominates;
but later on in the coarsening process the precipitates will be
of this size or larger, so that elastic effects become important.
As an example, the value of $R_0$ for
Ni$_3$Al precipitates in nickel is of order $10^{-7}$ m,
and it has been shown \cite{95par2} 
\ds by experiments on a sequence of alloys with different values for $R_0$, 
that some of the effects due to elastic misfit are proportional to $R/R_0$ 
at small $R/R_0$.

We shall describe three different types of model which have been
used in the theory of these elastic effects.
In the first, the two phases are treated as continuous materials
obeying the macroscopic laws of linear elasticity theory
\footnote{Grinfeld \cite{86gri} is alone in taking seriously the
possibility of nonlinear effects},
and the interface between them is treated as a geometrical surface
({\em i.e.} its thickness is ignored). This was the model used
by J. D. Eshelby in his pioneering work 
\cite{51esh}, \cite{57esh}, \cite{61esh}.
We shall call it the sharp interface model. There 
are two versions of this model : a `static' version in 
which the energies of different arrangements of the phases are
compared without detailed consideration of how the system
can get from one arrangement to the other, and a `dynamic'
version in which the mechanism for such changes, namely
diffusion, is also included in the model.
The static version will be considered in chapter \ref{SSImodel} and
the dynamic version in chapter \ref{SImodel2}. 

A second type of model, introduced by John Cahn in 1961,
takes the structure of the interface into account
by using the concentration of one of the alloy components
as a field variable. Such models are sometimes described
as `mesoscopic'. The concentration field is approximately
constant (but at different values) in the two phases, and it
varies continouously across the interface. Its time variation
is given by a deterministic differential equation, analogous to 
the Cahn-Hilliard equation \cite{61cah}. We shall call this the
diffuse interface model and describe it in chapter \ref{DImodel}.

In the third type of model a completely microscopic description 
is used. The atoms are no longer represented
by a continuum; instead we follow them
individually and model their diffusion
by random jumps. Although it gives no analytic results,
this method has the merit that the physical assumptions going
into it are very simple. It also takes fluctuations
into account and therefore includes the possibility of nucleation
in a natural way. We shall call it the atomic lattice model
and describe it in chapter \ref{ALM}.

In cases where the lattice parameters are approximately the
same in the two phases, so that the elastic misfit is
unimportant, all three of these methods have been found to give 
results in reasonable agreement with experiments. For reviews, see
Gunton\cite{83gun1}, Furukawa \cite{84fur}, \cite{85fur}, Binder\cite{91bin}.
However, when the misfit is important, the theoretical situation
becomes more complicated, and it may be that a combination of all
the three methods, or more, is necessary for a proper understanding.

Some effects on morphology and kinetics typically observed in
experiments are summarized in the next section. Many of these effects
are actually reproduced by the various theoretical approaches.
These effects range from a change (typically a slowing down) in the rate 
of precipitate coarsening
to a transformation from
nearly spherical to cube-like or plate-like precipitates aligned
along the elastically soft directions, that is, [100] and equivalent ones
in the case of Ni-base alloys.
%
%
%
One also observes a direct effect of external stresses on the coarsening
behaviour of alloys, which can lead to the formation of cylinders or
parallel plates with wavy interfaces                             

\subsection{Experimentally observed effects due to elastic misfit interactions}
\label{exp}

The aim of the present chapter is not to give an exhaustive review of
the effects of elastic misfit strain on the phase separation process
but rather to show some typical examples. Because of the huge
practical importance of nickel-base superalloys for use at high
temperatures, e.g. in turbine blades, these alloys have
attracted much interest in recent experimental studies. Most of
the following examples will focus on this type of alloy system where
the nickel-rich matrix ($\gamma$-phase) has a face-centered cubic
structure and contains $\gamma'$-precipitates with L$_1$2 structure of
the type Ni$_3$X, where X may be Al, Ti, Si, etc.

\subsubsection{Morphological effects} \lbl{morph}

Depending on the composition of the alloy, the
precipitates may be either round as, e.g. in Ni-Al-Si \cite{HaydnChen}
or cube-like as in, for example, Ni-Al \cite{93mah}. The cube-like shape is,
in fact, very frequent in this type of system, appearing also
in ternary alloys like Ni-Al-Mo
\cite{89con,90cal,95seq1,95seq2,95fah,97lab}, Ni-Al-Ti
\cite{89hei}, Ni-Al-Cr or Ni-Al-Si \cite{89miy}.
The generally accepted reason for the cube-like shape is the effect of
anisotropic elasticity. A further observation is
the alignment of these precipitates, like strings of pearls,
along the elastically soft directions.

As an example, we show in Fig. 1 typical transmission electron microscopy
(TEM) and small-angle X-ray scattering (SAXS) data for Ni-Al-Mo
alloys. This ternary alloy system has the property that the lattice misfit
depends on the molybdenum content in such a way that negative, positive or
zero misfit is possible \cite{89con}. Fig. 1a shows the TEM data for an
alloy with no misfit. The $\gamma'$ precipitates, which appear white, are
round in this case, and arranged at random.  The SAXS pattern (Fig. 1b),
which represents roughly the square of the Fourier transform of the
microstructure (as depicted in Fig1a), is completely isotropic. This
symmetry shows that there is no preferred direction in the configuration of
the particles. This situation is exactly what one expects for a precipitate
microstructure which coarsens to reduce the total amount of interface
between matrix and precipitates.

For comparison, Figs. 1c and 1d show the corresponding data for a
Ni-Al-Mo alloy with similar fraction of $\gamma'$ phase but where the
molybdenum content was adjusted to give a lattice misfit. Now the
precipitates are \ds cube-like rather than spherical and they are arranged
like strings of pearls along the cubic directions [100] or
equivalent ones. Correspondingly, \ds the SAXS patterns are no longer
spherically symmetric, although they still have cubic symmetry.
Fig. 1e and Fig. 1f show a similar situation but with a higher volume
fraction of $\gamma'$ precipitates.
Strongly periodic arrangements of precipitates are also found in
alloys like Ti-Ni \cite{90cer,97vys}.

In some other types of alloys, a typical example being Al-Cu
\cite{kostorz} it is also common to find plate-like
coherent precipitates as a result of elastic misfit strains. The
example of the so-called `Guinier-Preston zones' in Al-Cu is
particularly striking, since these coherent  precipitates are
practically monoatomic layers of copper which create large elastic
distortions in the aluminium matrix. Other examples of alloys with
plate-like precipitates are Cu-Be \cite{groger} and internally oxidized
Cu-Fe \cite{cufeo1,cufeo2}.

A striking effect has been observed in  
alloys of the type Co-Pt \cite{91ler} or (CuAu)-Pt \cite{95udo} where
precipitates with tetragonal structure are developing in a cubic matrix.
Fig. 2a (from \cite{98bou}) shows the tile-like structures 
that can develop.

\ds \ds Finally, it has been found in some nickel-base alloys that
large cube-like precipitates may even split into several smaller ones
\cite{94qu,84doi}
a process which has also been related to elastic interactions
\cite{88kha} (see Fig. 4).

\subsubsection{Effects on coarsening kinetics}

The reported effects on the rate of coarsening are seemingly less
consistent. Indeed, an increase of the mean precipitate radius
proportional to $t^{1/3}$, $t$ being the annealing time, has been
reported in some cases, like Ni-Al-Mo \cite{90cal,95fah}
or Ni-Al-Si \cite{HaydnChen}. This growth law is
very similar to what is usually found in alloys without lattice misfits
and in computer simulations. It is due to a ripening process, described in the
Lifshitz-Slyozov-Wagner (LSW) theory \cite{61lif,61wag}, where the large
precipitates grow at the expense of the small ones, thereby reducing
the total interface area between matrix and precipitates.

In some other cases, e.g. Ni-Cu-Si alloys, the coarsening starts out
according to a $t^{1/3}$ but slows down in later stages,
eventually even coming to a standstill. This behaviour is illustrated
in Fig. 3 where the growth of the mean precipitate size $R$ is plotted for
Ni-Al-Mo (Fig. 3a) and for Ni-Cu-Si (Fig. 3b) alloys.
Another example of this type are titanium-rich $\omega$-phase
precipitates in Ti-Mo \cite{94lan,91fra} (see Fig. 3c).
The usual interpretation of this slowing down is {\em inverse coarsening},
which means that, because of the elastic interaction, 
smaller precipitates may grow at the expense of larger ones even though this
increases the interfactial area.

There is, however, no generally accepted
explanation for the fact that slowing down of coarsening is observed
in some cases but not in all.
One possibility is that the inverse
coarsening may reduce the prefactor $\lambda$ in the expression
\be R \approx \lambda t^{1/3} \ee
instead of affecting the exponent 1/3. This is, indeed, observed in
some Ni-Al-Si
alloys \cite{HaydnChen}
Another proposal is that there could be an interplay
between elastic heterogeneity and anisotropy 
such that inverse coarsening
would only become predominant in cases where the anisotropy is not too
large \cite{95par3,96fra}.

\subsubsection{The role of applied stress; rafting} \lbl{raft}

When an external stress is applied to an alloy
it can cause the shapes and arrangement of the precipitates to change.
For example, as already mentioned and shown in Fig. 1c-f, in a
nickel-base
alloy annealed without external stress the precipitates
are approximately cubes and are arranged in a cubic array, in conformity
with the cubic symmetry of the crystal. If a tension
is applied along the (100) axis the precipitates shorten along this
axis and widen along the other two axes. At large enough tension
the precipitates join up and form plates at right angles to the (100)
direction.
On the other hand a compression along the (100) axis causes
the precipitates to lengthen along this axis and eventually form rods.
This behaviour is shown for the case of the Ni-Al-Mo alloy system in Fig. 1g.
The corresponding SAXS pattern (Fig. 1h) shows clearly the breaking of
the cubic symmetry. Pictures very similar to Fig. 1h
have been published very recently for a uniaxially strained
commercial nickel-base superalloy studied by small-angle neutron
scattering \cite{97ver}.

This phenomenon, known as rafting, was first observed experimentally
by Webster and Sullivan \cite{67web}, Sullivan {\em et al.} \cite{68sul} and
Tien and Copley \cite{71tie1}. A well-written short review of
the subject is given in the paper by Nabarro {\em et al.} \cite{96nab}.
The mechanism of rafting is thought by many authors
(Socrate and Parks 1993 \cite{93soc}, 
Pollock and Argon 1994 \cite{94pol},
Vall\'{e}s and Arrell 1994 \cite{94arr,94val},
Buffi\`{e}re and Ingot 1995 \cite{95buf}, 
Svoboda and Luc\'{a}\v{c} \cite{96svo}
Ohasha {\em et al.} 1997 \cite{97oha}) 
to involve not only the diffusional and elastic effects considered here 
but also dislocations and plasticity; however
such considerations are beyond the scope of this paper.

\section{The static sharp interface model }\lbl{SSImodel}

In the sharp interface model the two phases, which we call $\alpha$ and
$\beta$, are modelled as distinct regions of space
which, together with the sharp interface between them,
make up the entire region $\Omega$ occupied by the
specimen. The two regions will be called
$\Omega^\alpha$ and $\Omega^\beta$, and the
interface between them will be denoted
by $\Gamma$. 
It frequently happens that
one of the phases consists of disconnected domains
called precipitates or inclusions, while the other phase,
called the matrix, occupies the remaining region
which looks something like a Swiss cheese.
In this case we shall normally
take $\Omega^\alpha$ to comprise the inclusions
and $\Omega^\beta$ to be the matrix.

\cs There are two versions of the sharp-interface model, which we shall
call the static and dynamic versions. 
In the static version, time evolution is not considered, and the
local composition of the alloy need not be considered either;
the main thing that the model can do is to compare the (free)
energies of various configurations of inclusions.
In the dynamic version, which we shall consider in the next
chapeter, we do consider time evolution, and since the mechanism
for time evolution is diffusion the
local composition of the alloy must also be included in the model.

\cs Although the static model contains no explicit time evolution
and therefore no coarsening, it does 
provide information about time evolution and coarsening through the
principle that the free energy must decrease with time.
For example, we can look 
for the energy-minimizing shape of an isolated inclusion of a given 
volume, and under suitable conditions the actual inclusions having this volume
should be close to this shape (which is not necessarily spherical,
even in an isotropic material). 
The theory of the model also contains important general results
about the elastic energy, such as the `Bitter-Crum' theorem
(see section \ref{theorems}), and the elasticity theory
contained in it underpins the models described in later sections.

\subsection{Theory of the energy of inclusions in an elastic medium} 
\lbl{inclusions}

The free energy of the system can be written in the form
\be
F = \int_\Omega (f + w)\dd^3\bx + F^\Gamma + W^{ext}
\lbl{free-energy}
\ee
where $f$ denotes the thermodynamic free energy density  at zero stress
$w$ is the elastic free energy density 
(defined by making $f+w$ the total free energy density),
$F^\Gamma$ denotes the 
free energy of the interface
and $W^{ext}$ denotes the potential energy of any
external mechanism that the specimen may
be connected to, for example a weight attached to one end of the specimen.
\ds \ds In the static version of the model, which we are considering 
in this chapter, the value of $f$ can be different in the
two phases, but is uniform throughout each phase.
The variable of integration $\bx$ is the coordinate of the
relevant material point in the unstrained reference state of the
material, so that a given  material point always has the same 
value of $\bx$, regardless of where the deformation has
taken it to.

To write down a formula for the elastic energy density $w$
we first define the {\em displacement field} $\bu$; its
definition is that the material point $\bx$ in the
undeformed material\footnote{
There is some arbitrariness in the
choice of an `undeformed' lattice, since a uniform expansion
of an undeformed lattice gives another undeformed lattice.
Different choices of undeformed lattice are related by
a dilatation of the coordinate system, which adds the same constant multiple
of the unit tensor to $\partial u_i/\partial x_j$ and to the stress-free
strain tensors. 
This transformation leaves the
energy expression (\ref{w}) invariant or (if the externally applied
stress, modelled as in eqn (\ref{Wext1}), includes an isotropic component)
merely adds to it a constant which is independent of the state of the specimen.
Physically measurable quantities such as the energy differences between
different configurations
and the differences of the stress-free strains between the two
phases are not affected by the choice of undeformed lattice.
}
will be found at the point $\bx + \bu(\bx)$
in the deformed material. The {\em strain tensor}
is then defined (we are using linear
elasticity theory; for the nonlinear formula see
eqn (1,3) of \cite{59lan}) by giving its Cartesian
components\footnote{\cs Some authors define
$e_{ij}$ for $i\neq j$ without the factor 1/2 in eqn (\ref{strain}),
but if that is done $e_{ij}$ is not a tensor.}
\be
{e}_{ij}(\bx) = \half \left( {\partial u_i \over \partial x_j} +
                        {\partial u_j \over \partial x_i} \right)
\lbl{strain}
\ee
We shall also need the {\em stress-free strain} tensor
\footnote{This tensor has also been called the `intrinsic strain' 
\cite{85roi}, and the `spontaneous deformation' \cite{86roi}.}
${e}^0_{ij}(\bx)$, defined as the value which
the strain tensor at $\bx$ would have if the
material were uniform and unstressed. The stress-free strain
is taken to be a different constant in each of the two phases,
corresponding to their different chemical compositions.
The difference between the stress-free strains
in the two phases is called the {\em transformation strain} \ds \ds 
or {\em misfit strain}.
A common assumption is that, under zero stress,
one of the two phases is simply a scaled-up version of the other.
Let us denote the scale factor by $1+q$, so
that the stress-free lattice spacing of the inclusion material
is greater than that of the matrix material by a factor $1+q$,
corresponding to a relative difference in specific volume
of approximately $3q$ if $q$ is small.
Then the stress-free strain can be given the value zero in the matrix
(this corresponds to a particular choice for the reference
state of the material) and to have the simple form
\be
{e}_{ij}^0 = q \delta_{ij}
\lbl{iso-SFS}
\ee
in the inclusions.

In linear elasticity the elastic energy is a quadratic function
of the strain tensor. If we adopt the convention that the elastic
energy is zero at zero stress, we can write this quadratic function
\be
w = \half \sum_{ijmn} \lambda_{ijmn}
({e}_{ij}(\bx) - {e}^0_{ij}(\bx))
({e}_{mn}(\bx) - {e}^0_{mn}(\bx))
\lbl{w}
\ee
Here $\lambda_{ijmn}$, a fourth-rank tensor, is called
the (isothermal) {\em elasticity tensor} or {\em stiffness tensor}.
This tensor is positive definite, in the
sense that $w$ is positive unless the tensors ${e}_{ij}$ and ${e}^0_{ij}$
are equal. In general the components $\lambda_{ijmn}$ may vary with 
position (and with chemical composition, temperature, etc),
but in the sharp-interface model they are taken to
be constant in each phase, though not necessarily the same constant.

The  $\lambda_{ijmn}$'s are symmetric under interchange of $i$ and $j$, 
of $m$ and $n$, and of the pair $(i,j)$ with the pair $(m,n)$;
and they usually have further symmetries reflecting the symmetries of
the crystal lattice. 
The most symmetrical case of all is an
isotropic solid such as rubber, for which the elastic energy density
simplifies (in linear elasticity theory) to
\be
w = \half K\{\sum_k (\Delta{e})_{kk\}}\}^2 +
{G} \sum_{ij}\{(\Delta{e})_{ij} - {1 \over 3}\delta_{ij}\sum_k
(\Delta{e})_{kk}\}^2 
\lbl{iso-w}
\ee
where $K$ is the {\em bulk modulus},
$(\Delta{e}) _{ij}$ means ${e}_{ij} - {e}^0_{ij}$
and ${G}$ is the {\em shear modulus} (often denoted by $\mu$).
The non-zero elements of the elasticity tensor
in this case are 
$\lambda_{1111}= K + {4 \over 3} {G},
\lambda_{1122} =  K - {2 \over 3}{G}, \lambda_{2323} = {G}$ 
and the ones
related to these by symmetry, e.g. $\lambda_{2332} = \lambda_{1212}$.
A more realistic case for metallurgists is the cubic crystal, for which
in the conventional notation \cs
\bea
w &=& \half C_{11} \sum_i(\Delta{e})_{ii}^2 +
C_{12}\{(\Delta{e})_{11}(\Delta{e})_{22}
     + (\Delta{e})_{22}(\Delta{e})_{33}
     + (\Delta{e})_{33}(\Delta{e})_{11}\}
\nonumber\\
&+&
2 C_{44}((\Delta{e}) _{12}^2
+ (\Delta{e}) _{23}^2 + (\Delta{e}) _{31}^2)
\lbl{cubic-w}
\eea
so that $\lambda_{1111} = C_{11},
\lambda_{1122} = C_{12}, \lambda_{2323} = C_{44}$, etc. 

The formula (\ref{w}), which generalizes a standard
formula of elasticity theory (eqn (10,1) of \cite{59lan})
to the case of non-vanishing stress-free strains,
is due to Khachaturyan \cite{66kha}
and (independently) \ds Roitburd \cite{67roi}. 
The formula corresponds to a particular way of apportioning
the total free energy density between the two terms $f$ and
$w$, in which $f$ is defined to be the free energy at zero
stress. The alternative convention, to define $f$ as the
free energy at zero strain, has also been used
\cite{90nis}.

For the interfacial energy term $F^\Gamma$ in (\ref{free-energy})
we shall use the simplest reasonable assumption,
\be
F^\Gamma = \sigma |\Gamma|
\lbl{Fint}
\ee
where $\sigma$ is the surface tension, assumed constant,
and $|\Gamma| = \int_{\Gamma}\dd^2 \bx$
denotes the area of the {\em undeformed} interface.
More general possibilities have also been considered ---
anisotropic surface energy by Cahn and Hoffman (1974) \cite{74cah},
surface energy depending on the deformation as well as the undeformed
area by Gurtin and Murdoch (1975) \cite{75gur},
Cahn and Larch\'{e} (1982) \cite{82cah} --- but these
refinements have not been important in the theory of coarsening.

To model the externally applied forces in a simple way,
we shall assume that there are no body forces,
and that the tractions (forces per unit area)
applied to the surface $\partial \Omega$ of the specimen are
of the form $\sum_j t_{ij}^{ext}n_j$,
where $n_j$ is the unit outward normal vector
and $t_{ij}^{ext}$, which we assume to be symmetric and 
independent of position, is the {\em externally applied stress} tensor.
The energy $W^{ext}$ of the mechanism providing these tractions can
then be equated to its energy when the specimen is 
in the reference state minus the work it does in bringing
the specimen from its reference state to the state with
displacement field $u_i$, so that
\be
W^{ext} = {\rm const.} 
- \int_{\partial \Omega} \sum_{ij} u_i t^{ext}_{ij} n_j \dd^2 \bx \nonumber \\
\lbl{Wext1}
\ee
Choosing the constant to be zero, and using first the divergence theorem
(applied to the vector field obtained by multiplying $u_i$ by
a unit vector in the $j$ direction)
and then the formula (\ref{strain}) and the symmetry of $t^{ext}_{ij}$
we can write this in the alternative form
\bea
W^{ext} & = &
- \int_\Omega \sum_{ij} t^{ext}_{ij}{\partial u_i \over \partial x_j} \dd^3\bx
\nonumber \\
& = & - \int_\Omega \sum_{ij} t^{ext}_{ij}e_{ij} \dd^3 \bx.
\lbl{Wext2}
\eea
In this formula, $e_{ij}$ is the strain in the presence of the 
externally applied stress, which in linear elasticity is an affine function 
\footnote {That is, a constant plus a linear function}
of $t_{ij}^{ext}$. With neglect of terms quadratic in  the 
externally applied stress, therefore, the strain $e_{ij}$ in
(\ref{Wext2}) can be given the value it has when there is no
externally applied stress.

\subsubsection{Elastic equilibrium}

To evaluate the elastic terms in  (\ref{free-energy}),
we need the displacement $u_i$ at each point in space.
The condition determining
$u_i$ is that it should minimize the elastic energy
$\int_\Omega w \dd^3\bx + W^{ext}$.
The Euler-Lagrange equation for this minimization,
called the {\em equation of elastic equilibrium}
(eqn (2,6) of \cite{59lan}), is (since $t^{ext}_{ij}$ is a constant)
\be
\sum_j {\partial t_{ij} \over \partial x_j} = 0
\lbl{eleqm}
\ee
where $t_{ij}$ is the stress tensor, defined (eqn (3,6) of \cite{59lan}) by
\be
t_{ij} ={\partial w \over \partial u_{i,j}}
\lbl{stress}
\ee
and $u_{i,j}$ means $\partial u_i/\partial x_j$.

Using (\ref{w}) in (\ref{stress}) we find with the help of the symmetry
relation 
$\lambda_{ijmn} = \lambda_{jimn}$ that the stress tensor is related
to the strain tensor by
\be
t_{ij} = \sum_{mn} \lambda_{ijmn}({e}_{mn} - {e}^0_{mn})
\lbl{Hooke}
\ee
(Hooke's law) and is therefore symmetric. In the isotropic case,
with $w$ given by (\ref{iso-w}), Hooke's law takes the form

\be
t_{ij} = K \delta_{ij} \sum_k (\Delta e)_{kk}
+ 2G\{(\Delta e)_{ij} - {1 \over 3}\delta_{ij}\sum_k (\Delta e)_{kk} \}
\lbl{iso-Hooke}
\ee

To complete the specification of the elastic equilibrium problem,
we need conditions on the elastic field variables at the interface $\Gamma$.
and at the boundary of the specimen. On $\Gamma$,
the condition of coherency requires $\bu$ to be continuous:
\be
[\bu] = 0
\lbl{ucont}
\ee
where $[\bu]$ means $[\bu]^\alpha_\beta$, that is
$\bu^\alpha- \bu^\beta$, the difference between the
limiting values of $\bu$ on the two sides of the interface.
Moreover, the values taken by the field $\bu$ at the interface
(which describe the position of the interface in the deformed material)
must satisfy an energy minimization condition.
Since we are assuming that the energy of the interface depends
only on its position relative to the lattice, the surface energy
is not changed by changes of the value of $\bu$ at the interface,
and the condition of minimum elastic energy turns out to be
\be
\sum_j [t_{ij}]n_j = 0
\lbl{tcont}
\ee
where $[t_{ij}]$ means $t_{ij}^\alpha - t_{ij}^\beta$,
and $\bn$ is a vector normal to the interface; the physical
interpretation of eqn (\ref{tcont}) is that the tractions
exerted on the interface by the material on each side of 
it must be equal and opposite
\footnote{
If the interfacial energy depends on the deformations as well,
then an extra term appears in (\ref{tcont}); see Gurtin and
Murdoch (1975) \cite{75gur}, Cahn and Larch\'{e} (1982) \cite{82cah}  .
}.

The boundary conditions at the surface of the specimen
depend on the physical situation there.
If the boundary of the specimen is clamped, then
the system of equations (\ref{eleqm}) and (\ref{Hooke})
should be solved subject to
prescribed values of $u_i$ on $\partial \Omega$. In the more
common situation where known forces are applied at the boundary
the boundary condition (obtained by minimizing $F$ without
restricting $u_i$ at  the boundary) is that the surface
tractions due to the internal stresses
are equal to the externally applied surface tractions :
\be
\sum_j t_{ij} n_j = \sum_j t_{ij}^{ext} n_j
\lbl{BC}
\ee
where, as before, $n_j$ denotes a unit vector normal to the boundary
$\partial \Omega$.

It is also possible to use periodic boundary conditions, taking
$\Omega$ to be a cube (or rectangular prism) and requiring
the stress and strain fields to be periodic with unit cell $\Omega$.
In general, the displacement field  $u_i$ will not be periodic,
but by virtue of (\ref{strain}) it can be written in the form
\be
u_i = u'_i + \sum_j U_{ij} x_j
\lbl{u'}
\ee
where $u'_i$ is periodic
\footnote{Khachaturyan \cite{83kha} calls $u'_i$ the `local displacement'}
and $U_{ij}$ is a constant tensor, not
necessarily symmetric. Then, using (\ref{strain}),
the strain can be written in the manifestly periodic form
\be
e_{ij} = \bar{e}_{ij} + \half \left( {\partial u'_i \over \partial x_j} +
                        {\partial u'_j \over \partial x_i} \right)
\lbl{eu'}
\ee
where $\bar{e}_{ij}$ 
is the space-averaged strain tensor
\be
\bar{e}_{ij} = {1 \over |\Omega|}\int_\Omega e_{ij}(\bx) \dd^3 \bx
= \half(U_{ij} + U_{ji})
\lbl{bare}
\ee
The value of $\bar{e}_{ij}$ does not appear in the elastic equilibrium
equations but it can be determined by direct minimization of $F$ 
with respect to $\bar{e}_{ij}$; the minimum occurs when
\be
\int_\Omega t_{ij} \dd^3 \bx = \int_\Omega t_{ij}^{ext} \dd^3 \bx
\lbl{tmin}
\ee

\subsubsection{The interface condition for inclusions of fixed volume}
\lbl{interfcond}

The position of the surface $\Gamma$ (relative to the undeformed
material) is determined by the way the
atoms diffuse through the specimen. As they diffuse, the position
of $\Gamma$ will change, but the total volume of each of the 
regions $\Omega^\alpha$ and $\Omega^\beta$ will remain the same.
In the final equilibrium configuration, the position of $\Gamma$ 
will minimize the energy subject to given values of these two volumes.
In addition, if the various inclusions are far enough apart, the
volumes of the individual inclusions will change only slowly, 
and it is a reasonable medium-term approximation to minimize the energy 
with respect to $\Gamma$ keeping the energy of each separate
inclusion fixed (assuming, of course that such a minimum exists).
In this section we formulate a necessary condition which $\Gamma$ must
satisfy at any such energy minimum.

Let $\Gamma$ be the surface of an inclusion
$\Omega^\alpha$ in a matrix $\Omega^\beta$ and consider a
small displacement of
$\Gamma$ to a new position $\Gamma'$, each point $\bx$ on $\Gamma$
moving to point $\bx + \bn(\bx) \delta x(\bx)$ on $\Gamma'$ (see Fig. 5).
Here $\bn(\bx)$ is a unit vector perpendicular
to $\Gamma$ at $\bx$, pointing outwards from the inclusion
({\em i.e.} from $\Omega^\alpha$ to $\Omega^\beta$, and
$\delta x(\bx)$, a small scalar, is the signed perpendicular distance
from $\bx$ to the nearest point on $\Gamma'$, positive if $\Gamma'$
is in the $\beta$ phase of the $\Gamma$ configuration.
Denote the region between $\Gamma$ and $\Gamma'$
by $\delta \Omega^\alpha$. Since we are not considering diffusion at present,
the volume of the inclusion is fixed, so that the volume of $\delta
\Omega^\alpha$, 
with an appropriate sign convention, is zero. Thus we have
(to lowest order in $\delta x$)
\be
\int_\Gamma \delta x \dd^2 \bx = 0
\lbl{constvol}
\ee
Now consider the change in energy on going from $\Gamma$ to $\Gamma'$.
There is no change in the total thermodynamic free energy $\int f \dd^3 \bx$,
since the volume of the inclusion does not change.
The change (increase) in the elastic energy between the two configurations,
as given by eqn (\ref{free-energy}), is the sum of three terms:
\begin{enumerate}
\item the change in surface energy due to the change in the area of $\Gamma$.
This energy increase is equal (in three dimensions) to
the surface tension $\sigma$ times the increase in undeformed area
$\int_\Gamma (2 \kappa \delta x) \dd^2 \bx$, where $\kappa$ is the mean
curvature, 
taken positive if the centre of curvature is on the $\alpha$ side of $\Gamma$.
\item the change in total elastic energy due to the fact that (for positive
$\delta x$) the region $\delta \Omega^\alpha$,
which was previously on the $\beta$ side of the
interface, is now on the $\alpha$ side.
To lowest order in $\delta x$ this increase is
$\int_\Gamma [w] \delta x \dd^2 \bx$
where $[w] = w^\alpha - w^\beta$ is the difference between
the elastic energy densities on the two sides of the interface at $\bx$.
\item The change in the elastic energy
due to changes $\delta \bu$ in the displacement field outside $\delta
\Omega^\alpha$. 
Denoting these displacements by $\delta \bu$
and using the definition (\ref{stress}) of $t_{ij}$
we can write the energy change, to lowest order, as
\bea
&&\int_\Omega \sum_{ij} t_{ij} \delta u_{i,j} \dd^3 \bx \nonumber \\
& = &\int_\Omega \sum_{ij}\{
(t_{ij} \delta u_i)_{,j}
- t_{ij,j} \delta u_i \} \dd^3 \bx
\eea
where the notation $ (\dots){,j}$ means $\partial(...)/\partial x_j$.
The second term in the integrand is zero because of the
elastic equilibrium condition (\ref{eleqm}).  
Applying the divergence theorem separately to the
regions $\Omega^\alpha$ and $\Omega^\beta$, and assuming 
that the surface of the specimen is either held fixed,
so that $\delta \bu = 0$ there, or is too far away to matter,
we can simplify the remaining term to
\bea
\int_\Gamma \sum_{ij} n_j [t_{ij} \delta u_i] \dd^2\bx 
& = &\int_\Gamma \sum_{ij} n_j t_{ij}^\alpha [\delta u_i] \dd^2\bx
\eea
by (\ref{tcont}).
The last equation would also be true with $t_{ij}^\beta$
replacing $t_{ij}^\alpha$.
To evaluate $[\delta u_i]$ (which means $[\delta u_i]_\beta^\alpha$)
we write it as a line integral across $\delta \Omega^\alpha$
\bea
[\delta u_i]  = [\delta u_i]_\beta^\alpha = - [\delta u_i]_\alpha^\beta & = &
\int_0^{\delta x} \sum_j {\partial (\delta u_i) \over \partial x_j} n_j \dd
z \nonumber \\ 
& = & - \int_0^{\delta x} \sum_j \delta {\partial u_i \over \partial x_j}
n_j \dd z \nonumber \\ 
& = & - \delta x \sum_j [u_{i,j}]n_j
\eea
\end{enumerate}

Adding together the three contributions we find for the change in the
total energy $W$
\be
\delta W = \int_\Gamma\{ 2\sigma \kappa + [w]
- \sum_{ik} T_i [u_{i,k}]n_k \} \delta x \dd^2 \bx
\lbl{deltaW}
\ee
where we have defined
$T_i = \sum_j t^\alpha_{ij}n_j = \sum_j t^\beta_{ij}n_j$, the
traction 
at the interface.

For $W$ to be a minimum against arbitrary changes of shape or
position of the inclusion which satisfy the constant-volume
condition (\ref{constvol}) it is necessary for the coefficient
of $\delta x$ in the integrand of (\ref{deltaW}) to satisfy
\be
2\sigma\kappa + [w] - \sum_{ik} T_i [u_{i,k}]n_k= p
\lbl{p}
\ee
where $p$ is a Lagrange multiplier, the same at all points on $\Gamma$,
which can be thought of as the excess pressure inside the inclusion.
If there is more than one inclusion, $p$ can be different for different
inclusions.

If the inclusion is not in equilibrium with respect to changes of
shape or position, then its surface will tend to move, since
the energy can be decreased by moving the interface towards the
$\alpha$ phase in the places where the left side is greater
than the right side, and towards the $\beta$ phase in the
places where the right side is greater. The mechanism for this
movement of the interface is diffusion, which we discuss
in chapter \ref{SImodel2}.

\subsubsection{The Fourier transform solution} \lbl{FTS}

If the elasticity tensor is independent of position
({\em i.e.} , for the sharp interface model, if it is the same in both phases)
we speak of {\em homogeneous elasticity}. In that case,
the equation of elastic equilibrium, (\ref{eleqm}),
can be solved by Fourier transforms for the case where
$\Omega$ is a rectangular box with periodic boundary conditions.
In this way the elastic energy of an arbitrary arrangement
of inclusions in such a box, or an arbitrary periodic
arrangement of inclusions in an infinite specimen,
can be expressed as a sum or integral in Fourier space.
Moreover, instead of being restricted to two different values,
one in $\Omega^\alpha$ and the other in $\Omega^\beta$,
the stress-free strain tensor $e_{ij}^0$ can vary with
position in any way.
This method of solution has been much exploited by Khachaturyan;
see his book \cite{83kha}.

We define the Fourier components of the displacement vector and
the stress-free strain tensor by
\bea
\btu(\bk) & = & \int_\Omega {\rm e}^{\imag\bk\cdot\bx} \bu'(\bx) \dd^3 \bx 
\lbl{FT}\\
\tilde{e}^0_{mn}(\bk) & = &
\int_\Omega {\rm e}^{\imag\bk\cdot\bx} {e}^0_{ij}(\bx) \dd^3\bx
\lbl{FTs}
\eea
where $u'_i$ is defined in (\ref{u'}).
Using Parseval's theorem and the symmetries such as
$\lambda_{ijmn} = \lambda_{jimn}$, the elastic energy formula
implied by (\ref{w}) and (\ref{Wext2}) can be written
\bea
W(e_{ij}(\cdot)) 
& = & \int_\Omega (w - \sum_{ij} t^{ext}_{ij} e_{ij})\dd^3 \bx \nonumber \\
& = &  {1 \over 2|\Omega|}\sum_{\bk \ne 0}
\sum_{ijmn}(\imag \tu^*_i(\bk) k_j - \te^0_{ij}(\bk)) \lambda_{ijmn}
(-\imag \tu^*_m(\bk) k_n - \te^0_{mn}(\bk)^*) + \nonumber\\
& + &
{|\Omega| \over 2}\sum_{ijmn} (\bar{e}_{ij} -\bar{e}_{ij}^0)
\lambda_{ijmn} (\bar{e}_{mn} -\bar{e}_{mn}^0)
- |\Omega|\sum_{ij} t^{ext}_{ij} \bar{e}_{ij}
\lbl{FTw}
\eea
where $|\Omega|$ means the volume of $\Omega$,
the star denotes a complex conjugate, the $\bk$-summation goes over
the reciprocal lattice of $\Omega$,  $\bar{e}_{ij}$ means the space
average of $e_{ij}(\bx)$ (see eqn (\ref{bare}))
and $\bar{e}_{ij}^0$ the space average of ${e}_{ij}^0$.
Minimizing the right side of (\ref{FTw}) with respect to $\bu(\bk)$
and $\bar{e}_{ij}$, we obtain for the actual elastic energy
(see  \cite{66kha,69kha1,83kha}) 
\be
W = {1 \over 2|\Omega|} \sum_{\bk \ne 0} \sum_{ijmn} 
\te_{ij}^0(\bk)^* \Psi_{ijmn}(\bk) \te^0_{mn}(\bk)
-|\Omega| \sum_{ij} t^{ext}_{ij} (\bar{e}_{ij}^0 + \half {e}_{ij}^{ext})
\lbl{khachw}
\ee
where 
\be
\Psi_{ijmn}(\bk) = \lambda_{ijmn} -
\sum_{pqrs}\lambda_{ijpq} k_p {Z}_{qr}(\bk) k_s \lambda_{rsmn}~~~(\bk \ne 0),
\lbl{Psi}
\ee
${Z}_{ij}(\bk)$ being the inverse of the matrix
\be
({Z}^{-1})_{ij} = \sum_{mn}k_m \lambda_{imnj} k_n,
\lbl{sG}
\ee
and $e^{ext}_{ij}$ means the solution of
$\sum_{mn} \lambda_{ijmn} e^{ext}_{mn} = t^{ext}_{mn}.$
Note that $\Psi_{ijmn}$ is homogeneous of degree zero in $\bk$ : 
it depends only on the direction of the vector $\bk$, not on its length. 
Also, $\Psi(\bk)$ is positive semi-definite for each value of $\bk$, 
in the sense that $\sum_{ijmn}e_{ij}^*\Psi_{ijmn}(\bk) e_{mn} \ge 0$
for any tensor $e_{ij}$; this property follows from the fact that this 
last sum is the value of the obviously positive corresponding term of the 
sum over $\bk$ in (\ref{FTw}) for a particular value of $\bu(\bk)$.

In an alloy where $e_{ij}^0(\bx)$ takes just two values, $e_{ij}^\alpha$
in $\Omega^\alpha$ and $e_{ij}^\beta$ in $\Omega^\beta$, the definition
(\ref{FTs}) gives
\be
\tilde{e}_{ij}^0(\bk) = [e_{ij}^0] \theta(\bk) ~~~(\bk \ne 0) 
\lbl{ek}
\ee
where $[e_{ij}^0]$ stands for the transformation strain 
$e_{ij}^\alpha - e_{ij}^\beta$ and $\theta(\bk)$ is defined by
\be
\theta(\bk) = \int_{\Omega^\alpha} {\rm e}^{\imag \bk \cdot \bx} \ddx.
\lbl{theta}
\ee
Substitution of (\ref{ek}) into the double integral in (\ref{khachw})
gives Khachaturyan's formula \cite{66kha,83kha}
\be
W - W^{ext} = \half \sum_{\bk \ne 0} B(\bk) |\theta(\bk)|^2
\lbl{khachB}
\ee
where we define
\be
B(\bk) = \sum_{ijmn} [e_{ij}^0] \Psi_{ijmn} [e_{mn}^0]
\lbl{B}
\ee
Like $\Psi_{ijmn}(\bk)$, the function $B$ 
depends only on the direction of the vector $\bk$, not on its length.

Using Parseval's theorem, the sum over $\bk$ in (\ref{khachw}) 
can be expressed as an integral over position space :
\be
W = \half\int\int_\Omega 
\sum_{ijmn}({e}_{ij}^0(\bx)-\bar{e}_{ij}^0)
H_{ijmn}(\bx - \by) 
({e}_{mn}^0(\by)-\bar{e}_{mn}^0) \dd^3 \bx \dd^3 \by + W^{ext}
\lbl{LRint}
\ee
where $H_{ijmn}$ stands for
$|\Omega|^{-1}\sum_{\bk \ne 0} \Psi_{ijmn}(\bk) e^{-i\bk\cdot\bx}$,
the inverse Fourier transform of $\Psi_{ijmn}$,
and $W^{ext}$ is equal to the sum involving
$t_{ij}^{ext}$ in (\ref{FTw}).
Eqn (\ref{LRint}) shows how, in the case of homogeneous elasticity,
the effect of elastic forces is equivalent to a certain
two-point interaction \cite{83kha,89onu2}.
It is a singular interaction, however : 
it generally includes a delta-function at
zero range, arising from the fact that the $\bk$-space
average of $\Psi_{ijmn}(\bk)$ is not zero. For anisotropic 
elasticity this interaction also decays very slowly at large distances,
because its Fourier transform depends only
on the direction of $\bk$ and is therefore 
discontinuous as $\bk \to 0$. In three dimensions, $H_{ijmn}(\bx - \by)$
decays like the cube of the distance $|\bx - \by|$,
for distances small compared to the size of $\Omega$.

In the special case where the material is
isotropic (see eqn (\ref{iso-w}))
as well as being elastically homogeneous
and where the stress-free strain is a pure dilatation as in
eqn (\ref{iso-SFS}), only the delta-function component of
$H_{ijmn}$ is present, and eqn (\ref{LRint}) simplifies to
\be
W = {18KG \over 3K +4G}\int_\Omega (q(\bx)-\bar{q})^2  \dd^3 \bx
+ W^{ext}
\lbl{iso-khachw}
\ee
where $q(\bx)$ is defined to be $q$ inside the inclusions and
0 outside, and $\bar{q}$ is the space average of $q(\bx)$.
The result (\ref{iso-khachw}), without the term $W^{ext}$,
is due to Cahn (1961) \cite{61cah}. 
He writes the expression in front
of the integral sign more neatly as
$2E/(1-\nu)$, where $E$ is Young's modulus and $\nu$ is
Poisson's ratio.\footnote{
Young's modulus and Poisson's ratio are related to the shear and bulk
moduli by $K = E/3(1-2\nu), G = E/2(1+\nu)$. See, for example, 
Gurtin \cite{72gur}.}

The question of what formula replaces (\ref{LRint})
if the elastic stiffness matrix is heterogeneous 
({\em i.e.} different in the two phases) was
considered by Onuki in 1989 \cite{89onu2}
for isotropic elasticity and in \cs 1990 \cite{90onu,90nis}
for a cubic crystal. Treating the inhomogeneity as a small
perturbation, he found that to first order of perturbation
the elastic energy could be expressed as the sum of three parts
\begin{enumerate}
\item{a long-range interaction due to the anisotropy}
\item{a dipolar interaction which is proportional to the 
external stress and to the amount of inhomogeneity}
\item{a non-quadratic `Eshelby' interaction 
proportional to the amount of inhomogeneity
}
\end{enumerate}
Khachaturyan {\em et al.} (1995) \cite{95kha} carried this type of analysis
further, considering higher orders of perturbation. They found that
in the lowest order of perturbation
the inhomogeneity is equivalent to a four-point interaction;
in the next order, to a six-point interaction, and so on.

\subsubsection{Two theorems}\lbl{theorems}

Theorem 1. (Eshelby 1961) \cite{61esh}). 
If the elasticity tensor $\lambda_{ijmn}$
and the external stress tensor $t^{ext}_{ij}$ are independent of position,
$W^{ext}$ depends only on the total volume of the inclusions and not
on their sizes, shapes, number or arrangement.

{\em Proof} By the conditions of the theorem, the tensor $e_{ij}^{ext}$
defined just after eqn (\ref{sG}) is independent of position.
Hence eqn (\ref{Wext2}) can be written
\bea
W^{ext} & = & -\sum_{ij} e_{ij}^{ext}
\int_\Omega \sum_{mn} \lambda_{ijmn}e_{mn} \ddx \nonumber\\
& = & - \sum_{ij} e_{ij}^{ext} 
\int_\Omega (t_{ij}+ \sum_{mn} \lambda_{ijmn}e_{mn}^0 ) \ddx
~\hbox{by}~(\ref{Hooke}) \nonumber \\
& = & - \sum_{ij} e_{ij}^{ext}
\left\{\int_\Omega t_{ij} \ddx + \sum_{mn}\lambda_{ijmn}    
(|\Omega^\alpha|e_{mn}^\alpha + |\Omega^\beta|e_{mn}^\beta)   \right\}
\eea
by (\ref{tmin}). This last expression depends on the inclusions
only through their total volumes1, QED.

Theorem 2. (The `Bitter-Crum' theorem)
\footnote{It was first pointed out by Bitter \cite{31bit} that in
in a linear elastically isotropic material there is no elastic
interaction between two inclusions of a certain type, and Crum
\cite{40cru} (also, independently, Robinson \cite{51rob}) showed how to
generalize this result to a wider class of inclusions. \cs See also
Eshelby 1957, 1959 \cite{57esh,59esh}.
\cs The name `Bitter-Crum' theorem was, apparently, devised by 
John Cahn \cite{84cah}.}
Under the conditions of theorem 1, if in addition the
elasticity tensor and the stress-free strain tensor are
isotropic, then for an infinite or periodic specimen the elastic
energy $W$ depends only on the total volume of the
inclusions, not on their sizes, shapes, positions etc.

{\em Proof :} For periodic boundary condition, the result 
follows from the formula (\ref{iso-khachw}), since the
integral is equal to $|\Omega^\alpha| (q^\alpha - \bar{q})^2 
+  |\Omega^\beta | (q^\beta  - \bar{q})^2 $ and is therefore
independent of how the inclusions are arranged. For an infinite system,
the integral in (\ref{iso-khachw}) is replaced by one over all space,
and since $\bar{q}$ is now equal to $q^\beta$, the integral is 
equal to $|\Omega^\alpha|(q^\alpha - q^\beta)^2$ and is again independent of 
how the inclusions are arranged, QED.

Both theorems, especially the first, are somewhat counter-intuitive.
The first theorem tells us that in order to explain rafting using 
linear elasticity theory we must either use a 
heterogeneous stiffness tensor or else have more than one type of inclusion
so that the stress-free strain can be different in 
different inclusions. The second tells us, 
\cs as John Cahn has always seen very clearly \cite{95cah}, that
\cs if there is to be an elastic interaction affecting the shape or
separation of 
inclusions then at least one of the following conditions must be violated: 

\noindent
\begin{tabular}{ll}
1. & Matrix and inclusion(s) have the same elastic stiffness tensor\\
2. & The strain characterizing the misfit is isotropic
({\it i.e.} purely dilatational), \\
3. & The elasticity  tensor is isotropic, \\
4. & The crystal has no boundaries ({\it i.e.} it is infinite, or periodic). \\
5. & The stress depends linearly on the displacement field
\end{tabular} \\
\noindent
As we shall see, 
different violations produce different effects on the coarsening behaviour.

\subsection{Energy calculations for isotropic elasticity}\lbl{iso-energy}

\ds \ds In the rest of this chapter we consider some applications 
of the static sharp interface model. 
In the present section we shall use the approximation
that both the stiffness tensor and the stress-free strain are isotropic.
Although real metals are always anisotropic, such calculations
are useful because they can be used to investigate the
effect of elastic heterogeneity ({\em i.e.} of violations of condition
1 of the Bitter-Crum theorem). Effects due to anisotropy will be
considered in the following section, \ref{aniso-energy}

\subsubsection{Elementary calculations with isotropic elasticity and misfit}
\lbl{elementary} 

\cs \cs \cs 
We consider first two types of inclusion whose elastic energies in an infinite 
matrix can be calculated quite easily, the plate (slab) and the sphere.
\cs To allow for the possibility of finite volume fractions,
in which case it is not obvious {\em a priori} which phase will constitute
the inclusions and which the matrix, we shall allow the stress-free
strain to differ from zero in both phases, being given by the
isotropic formula (\ref{iso-SFS}) but with different values for
$q$ in the two phases.
The object of the calculations is to find
the space-averaged energy per unit volume and also the 
space-averaged strain (which gives information
about the response to an externally applied stress).

\noindent{\em Plate-shaped inclusions}

Consider a configuration consisting of alternating slabs of the two phases,
the slabs of phase $\alpha$ having thickness $h^\alpha$, those of phase
$\beta$ having thickness $h^\beta$. Denote the normal to the faces of 
the slabs by $\bn$. 
The equations of elastic equilibrium have a solution in which the
strain in the $\alpha$ phase has the form
\be
e_{ij}^\alpha = Q^\alpha \delta_{ij} + P^\alpha (\delta_{ij} - 3 n_i n_j)
\lbl{eQP-plate}
\ee
where $P^\alpha$ and $Q^\alpha$ are constants, and similarly for the
$\beta$ phase. The total energy per unit volume of material
can be calculated from eqn (\ref{iso-w}); for an isotropic stress-free
strain, as given by eqn(\ref{iso-SFS}) generalized to allow for
non-zero values of $q$ in each of the two phases, it is
\be 
\bar{w} = 
\{ {9\over 2} K^\alpha (Q^\alpha - q^\alpha)^2
+ 6G^\alpha (P^\alpha)^2 \}\phi  +
\{ {9\over 2} K^\beta (Q^\beta - q^\beta)^2
+ 6G^\beta (P^\beta)^2 \}(1-\phi)
\lbl{wQP-plate}
\ee
where $\phi = h^\alpha/(h^\alpha + h^\beta)$ is the volume fraction of
phase $\alpha$.

The expression (\ref{wQP-plate}) is to be minimized subject to the coherence 
constraint (\ref{ucont}),
which because of (\ref{strain}) and (\ref{eQP-plate}) is now
\be
P^\alpha + Q^\alpha = P^\beta + Q^\beta
\ee
(This minimization automatically takes care of the stress continuity
condition (\ref{tcont})).
The minimum value, which gives the actual mean energy density of this 
system of inclusions, is
\be 
\bar{w} = {9 \phi(1-\phi)\over 2} K_* [q]^2
\lbl{wmin-plate}
\ee
where $K_*$ is defined by
\be
{1 \over K_*} = 
{1-\phi\over K^\alpha} + {\phi \over K^\beta} + 
{3\over4}\left\{ {1-\phi \over G^\alpha} + {\phi \over G^\beta} \right\}.
\lbl{Kstar}
\ee
In the limit where the volume fraction of the precipitates goes to zero,
(this includes the case of a single plate-like inclusion in an
infinite matrix)
the energy {\em per unit volume of precipitate} is
\be
\lim_{\phi \to 0} \bar{w}/\phi = 
{18 G^\alpha K^\alpha [q]^2 \over 3 K^\alpha + 4 G^\alpha}
\lbl{platew}
\ee

If there is a small externally applied stress, the resulting
energy can be found from the average strain at zero applied stress, using 
(\ref{Wext2}). The anisotropic (deviatoric) part of this average strain is
\be
\bar{e}_{ij} - {1 \over 3}\sum_k \bar{e}_{kk} = 
-(\phi P^\alpha + (1-\phi)P^\beta) = 
{3 \over 4}\left({1\over G^\alpha} - {1\over G^\beta}\right)K_* [q]
\lbl{plateraft}
\ee
We shall not need the isotropic part.

It is not hard to generalize these results to a cubic crystal, if
$\bn$ is in a convenient direction.
If $\bn$ is along one of the crystal axes, it is only necessary to replace
$K$ by $(C_{11} + 2C_{12})/3$ and $G$ by $(C_{11} - C_{12})/2$ in the
various formulas. If $n$ is along the [111] direction, 
$G$ should instead be replaced by $C_{44}$.

\noindent{\em Spherical inclusion}

Consider a spherical inclusion of $\alpha$ phase, of radius $R_1$, 
embedded in a larger sphere of $\beta$ phase having
radius $R_2$. Putting the centres of both spheres 
at the origin, we look for a solution in the form  
\be
\bu =
\left\{
\begin{array}{ll}
Q^\alpha \bx                          & (|\bx| < R_1)  \\
Q^\beta  \bx + P \nabla{1\over |\bx|} & (R_1| < \bx| < R_2)
\end{array} \right .
\ee
where $Q^\alpha, Q^\beta$ and $P$ are scalar constants.
Since $\bu$ is radial in both solutions,
the coherency condition (\ref{ucont}) can be satisfied by making
\be
Q^\alpha R_1 = Q^\beta R_1 - P/R_1^2
\lbl{match1}
\ee
     
Using (\ref{strain}) we find that the strain is
\be
e_{ij} =
\left\{
\begin{array}{ll}
Q^\alpha \delta_{ij}                                           
& (|\bx| < R_1)  \\
Q^\beta  \delta_{ij} + P(3x_i x_j -|\bx|^2\delta_{ij})/ |\bx|^5
& (R_1 < |\bx| < R_2)
\end{array} \right .
\ee
Hence, by (\ref{iso-w}), the elastic energy is
\bea
W & = & {4\pi R_1^3 \over 3}
{9K^\alpha \over 2} (Q^\alpha- q^\alpha)^2 
+  {4\pi (R_2^3 - R_1^3) \over 3}
{9K^\beta  \over 2} (Q^\beta - q^\beta)^2 + 
\int_{R_1}^{R_2} 6 G^\beta \left({P^\beta \over r^3}\right)^2 4\pi r^2 \dd r 
\nonumber \\
& = & {4\pi R_2^3 \over 3}
\left\{{9 K^\alpha \over 2} (Q^\alpha- q^\alpha)^2 \phi 
+ {9 K^\beta \over 2} (Q^\beta- q^\beta)^2 (1-\phi) 
+ 6 G^\beta (Q^\beta - Q^\alpha)^2 \phi(1-\phi)
\right\}
\lbl{W-sphere}
\eea
where $\phi = (R_1/R_2)^3$ is the volume fraction of phase $\alpha$, 
and we have used (\ref{match1}) to eliminate $P$ in the second line .

The expression (\ref{W-sphere}) is to be minimized with respect to
$Q^\alpha$ and $Q^\beta$. 
The minimum value, which gives the actual
space-averaged elastic energy per unit volume, can be written
\be 
\bar{w} = {9 \phi(1-\phi)\over 2} K_\bullet [q]^2
\lbl{wmin-sphere}
\ee
where $K_\bullet$ is defined by
\be
{1 \over K_\bullet} = 
{ 1-\phi \over K^\alpha} +
{\phi \over K^\beta} + {3 \over 4G^\beta}.
\lbl{Kbullet}
\ee
In the limit of zero $\phi$, which includes the case of a single
spherical inclusion in an infinite matrix, the energy per unit volume
of inclusion is
\be
\lim_{\phi \to 0} \bar{w}/\phi = 
{18 K^\alpha G^\beta [q]^2 \over 3K^\alpha + 4G^\beta }
\lbl{spherew}
\ee

For this geometry, the space average of the strain is purely isotropic,
and so the linear part of the energy of a spherical inclusion
due to an externally applied field is zero.

\noindent{\em Conclusions}

>From these elementary calculations we can draw some tentative conclusions  
about the effects of elastic inhomogeneity on the 
energetically preferred configurations of the precipitates,
provided that that elastic anisotropy does not mask these effects.

First of all, a comparison of eqns (\ref{spherew}) and (\ref{platew})
shows that if ${G}^\alpha > {G}^\beta$, that is if the
inclusion is more rigid than the matrix, then the
sphere has lower energy than the plate; on the other hand
if the inclusion is less rigid than the matrix, then the plate
configuration has the lower energy. That is to say,
a spherical inclusion is likely to be stable only if it is
stronger (more rigid) than the surrounding matrix. 
\cs If it is justified to use eqn (\ref{wmin-sphere}) 
to estimate the energy of a system of spherical inclusions even when 
$\phi$ is not vanishingly small then one can go even further : 
this formula, when compared 
with (\ref{wmin-plate}), suggests that a system of spherical inclusions
made from the more rigid material will always have a lower energy 
than the alternating plate configuration with the same $\phi$, so that 
in addition to destabilizing soft spherical inclusions
elastic heterogeneity tends to destabilize all plate-like inclusions. 

Secondly, eqn (\ref{plateraft}) shows that for an alloy where the
more rigid material also has a larger stress-free volume
(the correlation between size and strength we
all remember from the school playground) 
the space-averaged strain is a contraction in the direction 
perpendicular to the plate and an expansion in the directions
parallel to it. Consequently in such a material, a plate-shaped
inclusion will tend to align perpendicular to 
an externally applied uniaxial compressive stress, but parallel
to an externally applied tensile stress. 
\footnote{The formulation (\ref{raftsign}) is given in \cite{96nab}.
According to Thompson and Voorhees \cite{97tho},
this effect was predicted by Cahn soon after his 1961 work on spinodal
decomposition : see ref. \cite{64cah})}
In symbols
\bea
t^{ext}_{axial}[q][G] > 0 & \Rightarrow &
\hbox{inclusions~parallel~to~axis~of~applied~stress~}\nonumber\\
t^{ext}_{axial}[q][G] < 0 & \Rightarrow &
\hbox{inclusions~perpendicular~to~}\dots
\lbl{raftsign}
\eea
where $t^{ext}_{axial} = (2t_{33}^{ext} - t_{11}^{ext} -t_{22}^{ext})/3 $
is the axial component of the anisotropic part of the applied stress
(taking this axis to be along the 3 axis), which is positive (negative) for
a tensile (compressive) stress.

These conclusions can be understood
physically in the following way: in the case of a spherical inclusion,
the matrix is sheared but the inclusion is not, so the sphere
is a good (low-energy) configuration when the \cs matrix is very flexible.
On the other hand, for a plate-shaped inclusion, only the inclusion
is sheared, and so when the inclusion is very flexible 
the plate is a better shape than a sphere. Moreover if, in addition to being 
softer than the matrix, the plate has a lower stress-free density,
then the matrix will prevent it from contracting in the direction
parallel to the plate; it can only contract in the direction perpendicular
to the plate, and hence the anisotropic part of the strain 
is a contraction in this direction, and 
plate-like inclusions will be energetically favoured when the 
stress is compressive. 

\subsubsection{The energy of a single inclusion : isotropic elasticity}
\lbl{iso-1}

For most shapes of inclusion other than sphere, cylinder and plate the
calculation of elastic fields is complicated, but Eshelby
\cite{57esh,59esh,61esh} showed how to do it for an ellipsoidal inclusion.
He found that if the inclusion is more (less) rigid
than the matrix, an ellipsoid has more (less) elastic energy than a 
sphere of the same volume. Thus, a spherical inclusion which is more rigid
than the matrix is stable against deformation to an ellipsoid,
but (\cs disregarding the stabilizing effect of the surface energy)
one that is is softer than the matrix is unstable.
This result is consistent with the comparison 
in the previous section between a spherical
inclusion and a plate, which can be regarded as an infinitely
flat ellipsoid.

Very few authors have used the condition (\ref{p}) for the
equilibrium shape of an inclusion, but Roitburd \cite{86roi} 
has shown that in the case where the transformation
strain is not a pure dilatation (thus violating condition 2 of the
Bitter-Crum theorem as well as condition 1), 
and where the surface energy is neglected, an ellipsoidal inclusion
satisfies this condition over a certain range of values of the parameters.

If we take surface energy into account as well as elastic energy,
the energy per unit volume of inclusion 
depends on the size of the inclusion as well as its shape.
Johnson and Cahn (1984) \cite{84joh2} showed that for small inclusions
the surface term $F^\Gamma$ in the free energy formula (\ref{free-energy})
predominates and the shape 
of the lowest-energy inclusion has the same symmetry as that which minimizes
$F^\Gamma$ for a fixed volume of the inclusion. In the case of an
isotropic surface tension this would be a sphere, but for anisotropic
surface tension it could also be \ds
a polyhedron facetted parallel to the crystal planes where the surface
tension is minimum. As the size is increased the 
elastic energy  becomes more important and there 
is a transition, which may be either continuous or discontinuous,
to a new equilibrium shape of lower symmetry.
If an external stress is present, this too can induce a shape transition
(Berkenpas {\em et al.} (1986) \cite{86ber}, 
Johnson {\em et al.} (1988) \cite{88joh}). 

With a view to constructing a theory of rafting, Pineau (1976) \cite{76pin}
calculated the energies of single elliptical inclusions in an externally 
applied stress field, using methods devised by Eshelby \cite{61esh}. 
A brief comparison of his predictions with experiment
is given by Socrate and Parks (1993) \cite{93soc}.

\subsubsection{The energy of two or more inclusions : isotropic elasticity}
\lbl{iso-energy2}

In the absence of elastic forces, the mechanism of coarsening
arises from the fact that a pair of spherical inclusions
with given total volume have less total surface energy the more unequal
their sizes. To see how this mechanism is influenced by
elastic forces, we would like to know how the elastic energy
of a pair of neighbouring inclusions depends on their size
and separation. Can they reduce their energy
by transferring material from one inclusion to another,
or by changing their separation?

For isotropic elasticity, calculations of this type were first done 
by Eshelby in 1966\cite{66esh}. For a pair of spherical inclusions with
radii $R_1,R_2$ and centres a distance $D$ apart, he found
the energy of interaction to be approximately
\be
W_{int} = {8 \pi \over 81} \left( 1 + \nu \over 1 - \nu \right) ^2
[q]^2 [G] \left\{
{R_1^6 R_2^3 \over (D^2 - R_2^2)^3} + {R_2^6 R_1^3 \over (D^2 - R_1 ^2)^3}
+ O([G])^2
\right\}
\lbl{EshelbyE}
\ee
where $\nu = (3K-{G})/(6K+2{G})$ is Poisson's ratio and $[q], [G]$ have 
the same meaning as in eqn (\ref{plateraft}). 
If the inclusions are more rigid than the matrix, then
formula (\ref{EshelbyE}) says that
the interaction energy is positive 
and (for fixed values of the separation $D$ and the total volume
$(4\pi/3)(R_1^3 + R_2^3)$) is least when
the two inclusions are of equal size. This suggests that
the elastic energy will tend to make the inclusions of equal
size. This stabilizing effect, just the opposite of the tendency of surface
energy to enlarge the large inclusions at the expense of the
small ones, has been given the name `inverse coarsening'

In his papers drawing attention to the above stabilizing effect
Johnson \cite{83joh1,84joh1} used a generalized version  of the formula
(\ref{EshelbyE}) which included also surface energy and the effect of an 
external stress. He concluded\cite{83joh1} that, in the absence of external 
stress, the inverse coarsening effect would occur whenever the size of the
inclusions exceeded a certain threshold related to the critical size
we have called $R_0$. He also found \cite{84joh1} 
that a uniaxial
external tensile stress could favour or inhibit inverse coarsening 
depending on the angle between the separation of the
inclusions and the axis of the external stress.

The elastic interaction of more complicated systems of inclusions
of given shape has also been studied; 
for example Johnson and Voorhees (1987) \cite{87joh2}
consider the elastic interaction energy of a system of several cuboids
(rectangular prisms), finding a tendency of such
precipitates to align, even though the medium is isotropic.
Abinandanan and Johnson (1993)\cite{93abi1,93abi2} calculated the elastic
interaction energy of two spherical inclusions with 
a tetragonal misfit strain. The energy is positive when the line joining the
particles and the axis of the misfit are parallel, negative 
when they are perpendicular.  

The value of all such conclusions based on 
the assumption of spherical precipitates is, 
however, called into question by the work of Onuki
and Nishimori (1991) \cite{91onu4}. They showed that the interaction
energy is extremely sensitive to the shapes of the inclusions.
Eqn (\ref{EshelbyE}) assumes that the inclusions are spheres,
but a small deformation of the spheres reduces the interaction to zero.
Since there is no reason why the actual precipitates should be exact spheres,
formula (\ref{EshelbyE}) cannot be relied on.

\cs Two other calculations using pre-assigned shapes of inclusion 
are the energy calculations for a three-dimensional  array of 
inclusions in the shape of square
or rectangular plates by Perovic {\em et al.} \cite{79per}
and the elastic stress calculations for two-dimensional arrays of
square plates by Glatzel and Feller-Kriepmayer \cite{89gla}.

\subsection{Energy calculations for anisotropic elasticity}\lbl{aniso-energy}

In this section we use energy methods to investigate 
the effect of elastic anisotropy (usually with cubic symmetry)
on coarsening --- that is to say, the effect of violating
condition 3 rather than condition 1 
of the Bitter-Crum theorem. Except where otherwise stated, the
material is assumed to be elastically homogeneous.

\subsubsection{The energy of a single inclusion : anisotropic elasticity}
\lbl{aniso-shape}

Khachaturyan (1966) \cite{66kha,83kha} 
(see also Roitburd (1967) \cite{67roi}) 
showed how to calculate the elastic energy of an inclusion of arbitrary shape, 
using the Fourier transform formula (\ref{khachw}).
This energy is minimized by making the inclusion a plate (or an array
of parallel plates) with faces (the `habit planes')
perpendicular to a vector $\bn$ that minimizes the function
$B(\bn)$ defined in (\ref{B}). This configuration obviously satisfies
the interfacial equilibrium condition (\ref{p}) too.
The directions of the minimizing vectors are called
{\em elastically soft} directions.
For a cubic crystal such as tungsten which has
positive anisotropy, {\em i.e.} $C_{11} - C_{12} - 2C_{44} >0$, 
the elastically soft (minimizing) directions are $[111]$, $[11\bar{1}]$ etc.; 
for negative anisotropy, as in Al, Cu, Fe, Ni, they are 
$[100],[010]$ and $[001]$.

The elastic energy of some other shapes of inclusion can also be found 
analytically, for example spheres, ellipsoids \cite{69kha1,69kha2},
and cubes \cite{78lee,83kha}. 
If the strain in the inclusion is uniform 
({\em i.e.} for all these shapes except the cube), 
the energy can also be calculated for a material 
which is elastically heterogeneous as well as anisotropic
\cite{77lee} (or section 8.4 of \cite{83kha}). 

When surface energy is taken into account, the minimum-energy shape
of the inclusion is no longer a plate, but it has still been computed 
analytically in some cases.
The two-dimensional calculations of Wen {\em et al.} (1981)
\cite{81wen1,83kha}  
illustrate how the minimum-energy shape, round when the inclusion is
very small, can pass through a succession of more and more plate-like shapes 
as the inclusion is made larger. For a crystal with square symmetry,
as the size of the inclusion is increased there is a transition from
minimum-energy shapes with fourfold (square) symmetry 
to ones with twofold symmetry \cite{94tho,96su1}. 
\cs (Symmetry-breaking transitions of a similar kind occur
for isotropic elasticity : see section \ref{iso-1})

In three dimensions, the minimum-energy shape is more difficult to compute; 
a common procedure has been simply to compare the energies of
various easily calculable shapes such as spheres, ellipsoids, cubes,
cuboids, tetrahedra and octahedra and to assume that the actual shape
at a given volume of inclusion will be similar to the easily calculated
shape having the least energy. \ds \ds Such calculations 
\cite{88kha,89kau,89gla,92mcc2}
show that the precipitate may pass thorough a variety of different
minimum-energy shapes as its size is increased, and may even split 
into two or more smaller pieces.
As we have mentioned in section \ref{morph}, splitting of this kind 
has been observed experimentally 
with $\gamma'$ precipitates in nickel-base and other 
superalloys \cite{82miy,85doi,89kau,94qu} : see Fig. 4.
A review of all such energy calculations up to 1992 is given by
Johnson and Voorhees \cite{92joh4,92joh3}. \ds

The effect of an externally applied stress on the energy-minimizing 
shape of an ellipsoidal inclusion with elastic forces only, 
a key factor in the theory of rafting, was studied by
Chang and Allen (1991) \cite{91cha}, following earlier  theoretical
work by Tien and Copley (1971) \cite{71tie1},
Pineau (1976) \cite{76pin} (mentioned in section \ref{iso-1} above),
Miyazaki {\em et al.} 1979 \cite{79miy}, 
Johnson (1987) \cite{87joh} (who showed that the true equilibrium shape
could be approximated by an ellpsoid of revolution), and
Johnson {\em et al.} (1988)) \cite{88joh}. \cs \cs 
Using realistic elastic constants, Chang and Allen find good
agreement with observed shapes and orientations. 
Their article also contains a good critical review of the
earlier theoretical work on rafting.
Nabarro {\em et al.} (1996) \cite{96nab} consider a cubic crystal with 
inclusions having  the shape of tetragonal prisms instead of ellipsoids, 
and they find that eqn (\ref{raftsign}) holds, with the shear modulus $G$ 
replaced by its analogue for a cubic crystal, which is 
(see the paragraph after (\ref{plateraft})) $(C_{11} - C_{22})/2$
if [100] is a soft direction.

The restriction of working within a pre-selected set of 
shapes (spheres, cuboids, etc.)
is avoided in the work of McCormack {\em et al.} 1992 \cite{92mcc1} 
who do an unusual type of two-dimensional finite-element 
computation in which, as the size of the inclusion is increased, 
its shape follows a path of steepest descent with respect to energy, 
not even requiring the topology to stay the same.
As the size of the inclusion is increased, the round shape first
turns into a square, and later on a nucleus appears
inside the square (or on the surface if the calculation forbids
changes of topology) and the square splits into two or four.

\subsubsection{The energy of several inclusions : anisotropic elasticity} 
\lbl{aniso-pair}


As in the case of isotropic elasticity, we study pairs of
inclusions in order to understand when inverse coarsening
can take place, and if it does what relative position of the inclusions
gives the least energy. In addition, there is the possibility which we have
already mentioned (see Fig. 4) that a single inclusion may be able to 
reduce its energy by splitting into two (or more) parts.

Several authors have studied these questions by investigating
the energy of a pair of spherical inclusions. 
\hs Johnson and Lee (1979) \cite{79joh} show that two spherical
inclusions in an anisotropic matrix can reduce their energy 
if they come close together
and align themselves along the 'soft' crystal directions.
Miyazaki, Doi and others \cite{86miy,92doi,93miy} find in addition that 
inverse coarsening is possible if the particles are close together 
and the misfit strain and particle size are large enough.
Shneck {\em et al.} (1992) \cite{92shn} give details of the elastic fields for 
one and two spherical inclusions in a cubic crystal with negative
anisotropy.

Inclusions shaped like cubes and cuboids have
also been studied, because of their relevance to the observed splitting of
such inclusions in $\gamma'$ precipitates.
\ds Doi {\em et al.} (1984) \cite{84doi} showed that a single  cuboid may
have more  
elastic energy per unit volume of precipitate than an array of eight cuboids, 
or a pair of parallel plates. 
Khachaturyan {\em et al.} (1988) \cite{88kha} compare the energy of a
cube-shaped   
inclusion,  in a cubic crystal having negative anisotropy,
with that of two half-cubes and with an octet of smaller cubes. 
Depending on the distances involved, either the doublet or the octet
can have lower energy than the single cube.

\subsubsection{Positional correlations} \lbl{PC}

\ds We have already noted that a pair of inclusions 
of given shape often has the least energy if the inclusions are close 
together and are aligned along an elastically soft direction.
The same tendency was found in arrays consisting of a large number of
inclusions, by Khachaturyan 1969 \cite{69kha0,83kha}, 
and also by
Khachaturyan and Airapetyan 1974 \cite{74kha}. They found, for example, 
that in  a cubic alloy with negative anisotropy the minimum-energy
array for a system of spherical inclusions 
is a simple cubic superlattice. 

\section{The dynamic sharp-interface model} \lbl{SImodel2}

\subsection{Modelling diffusion}\lbl{MD}

\cs The dynamic version of the sharp-interface model differs from the
static version in taking account of the mechanism of time evolution.
which is diffusion. Since diffusion arises from concentration gradients,
we drop the assumption that was made in 
the static version of the sharp-interface model 
that the intrinsic properties of the alloy
are uniform in each phase. The composition of the alloy
will now depend on position; but the elastic stiffness matrix
and the stress-free strain tensor will still be assumed
uniform within each phase, though in general different in different phases.

\ds \ds For simplicity, let us assume that the alloy consists of
just two types of atom, $A$ and $B$, so that the local
composition of the alloy can be described by a single
field, the local concentration of $A$ atoms,
which we denote by $c$. We shall assume that the concentration
of $A$ atoms in the $\beta$ phase is small
so that $c$ is small in the $\beta$ phase. 
In the $\alpha$ phase, we assume that $c$ is
close to its zero-temperature equilibrium value in that phase 
which we denote by $c_0^\alpha$. 
If the $\alpha$ phase is pure $A$, then $c_0^\alpha = 1$,
but it is also possible for $c_0^\alpha$ to be less than 1; for
example in an Ni-Al alloy the precipitates have a Ni$_3$-Al
structure so that for this alloy the equilibrium concetration of Al 
atoms in the precipitates is $c_0^\alpha=1/4$ .

\subsubsection{The diffusion equation and its boundary conditions}\lbl{debc}

\cs \cs \cs 
Within each phase, the concentration varies with time by diffusion,
which we can model by the standard diffusion equation
\be
{\partial c \over \partial t} = \nabla\cdot(D\nabla c)
\lbl{diffneq}
\ee
For simplicity, we shall take the diffusivity $D$ to be uniform in each phase, 
though it may take different values in different phases.

The motion of the interface is controlled by the net rate of arrival
of $A$ atoms, in accordance with the standard mass conservation formula
associated with (\ref{diffneq}),
\be
v_n[c] = - [D \bn \cdot \nabla c]
\lbl{Gammadot}
\ee
where $\bn$ is a unit vector normal to $\Gamma$ and 
$v_n$ denotes the velocity of $\Gamma$ (with respect to the
undistorted lattice) in the direction of $\bn$.

To use these equations we need two conditions to determine the value of
$c$ on the two sides of $\Gamma$. Since these conditions are also
satisfied at equilibrium, we can obtain approximations 
to them by formulating the two
conditions that must be satisfied on $\Gamma$ at equilibrium.
One of these conditions refers to the possibility of mass transfer 
across $\Gamma$, the other to the possibility of motion of $\Gamma$ 
relative to the underlying lattice.

The condition for equilibrium under mass transfer across $\Gamma$
is that the chemical potentials of the two kinds of atoms 
on the two sides should be equal. 
Denoting the chemical potential of $A$ relative to $B$ 
({\em i.e.} the excess of the 
chemical potential of $A$ over that of $B$) by $\mu$, 
we can write this condition
\be
\mu^\alpha = \mu^\beta  ~~~\hbox{on}~\Gamma
\lbl{equalmu}
\ee
where $\mu^\alpha$ (or $\mu^\beta$) denotes the limiting value 
of $\mu$ at any point on $\Gamma$ as that point
is approached from within the $\alpha$ (or $\beta$) phase. 

If the system is not in equilibrium, 
extra terms may appear in (\ref{equalmu})
depending, for example, on the velocity
of the interface (corresponding to the kinetic undercooling
term in the theory of freezing and melting). We shall assume 
here, however, that the deviation from equilibrium is small enough to justify
neglecting such terms.

In defining the chemical potentials which appear in (\ref{equalmu}), 
we must take account of the
elastic contribution to the free energy, defining it as the
derivative of the total free energy with respect to particle number at
constant strain. However, since we are taking the stiffness
matrix and the stress-free strain to be independent of concentration
in each phase, it follows from the formula (\ref{w}) for the strain 
energy that the elastic contribution to the free energy does not
affect the chemical potential. The formula giving the 
chemical potential (of $A$ relative to $B$) is therefore
just the same as it would be in the absence of elastic energy:
\be
\mu(c) = {df(c) \over dc}
\lbl{sfmu}
\ee
(If temperature variations were of any importance, we would
of course write $(\partial f /\partial c)_T $, but in a metal 
it is reasonable to treat all processes as isothermal.)

To obtain our second condition, we consider the condition for equilibrium
under displacements of the interface. This can be obtained by 
requiring the total free energy to be a minimum with respect to
variations of $\Gamma$ subject to the constraint
that the number of $A$ particles, denoted here by $N^\alpha$, is fixed: 
\be
N^\alpha = \int_\Omega c \dd^3\bx = ~~~{\rm const.}
\lbl{c-cons}
\ee
The argument closely follows the corresponding one in 
section \ref{interfcond}, with the constant-volume constraint 
used there (see eqn (\ref{constvol})) replaced by (\ref{c-cons}),
and the resulting necessary condition for a minimum, 
analogous to (\ref{p}), is
\bea
2\sigma\kappa + [f+w] - \sum_{ik} T_i [u_{i,k}]n_k 
& = & \mu[c]
\lbl{robin}
\eea
where 
$[c]=c^\alpha-c^\beta$ is the discontinuity in $c$ across the interface.
We have denoted the Lagrange multiplier by $\mu$
in accordance with the general princple that
the Lagrange multiplier is equal to the
derivative of the minimum value with respect to the value of the constraint,
which in this case is $dF/dN^\alpha$, {\em i.e.} 
the chemical potential $\mu$. At equilibrium, $\mu$, $\mu^\alpha$ and
$\mu^\beta$ are all the same, and as in the case of eqn (\ref{equalmu})
we shall make the approximation of neglecting any non-equilibrium
terms in (\ref{robin}), proportional to, say, the velocity of $\Gamma$,  
even when the system is not in equilibrium.

The condition (\ref{robin}) (without the curvature term)
is due to Robin \cite{74rob} and, in greater generality,
to Larch\'{e} and Cahn (1978) \cite{78lar2}. The curvature
term, which incorporates the Gibbs-Thomson effect, was
included by Cahn and Larch\'{e} (1982) \cite{82cah}, following
earlier work by Gurtin and Murdoch (1975) \cite{75gur} on surface
stresses in solids.

\subsubsection{The generalized Gibbs-Thomson condition}

\cs \cs \cs Eqn (\ref{robin}) can be brought to a more convenient form,
generalizing the Gibbs-Thomson formula. 
For each phase we introduce
(following Larch\'{e} and Cahn 1973 \cite{73lar})
a thermodynamic potential $\pi$, the grand
canonical pressure, referring to the properties of this
phase at zero stress. Considered as a function of $c$, it is defined by
\be
\pi(c) = c \mu(c) - f(c)
\lbl{pi}
\ee
so that
\be
{d \pi(c) \over dc} = c {d \mu \over dc}.
\lbl{pi'}
\ee
The interface condition (\ref{robin}) can now be written

\be
[\pi] = 2\sigma\kappa + [w]- \sum_{ik} T_i [u_{i,k}]n_k
\lbl{pikap}
\ee
where $[\pi]$ means $\pi^\alpha - \pi^\beta$, that is
$\pi(c^\alpha) - \pi(c^\beta)$.

To obtain convenient approximate formulas for 
$\mu$ and $\pi$, consider first their values at the type of
equilibrium normally considered in thermodynamics, for which 
the interfacial curvature is small and the lattices in the
two phases are incoherent, so that surface and elastic
energies play no part in the equilibrium conditions.
At such an equilibrium eqn (\ref{equalmu}) shows that 
there will be a common value of 
$\mu^\alpha$ and $\mu^\beta$; denote this common value by $\mu_{eq}$.
Moreover, eqn (\ref{pikap}) reduces in this case 
to $[\pi] = 0$; that is to say, $\pi^\alpha$
and $\pi^\beta$ have a common value at such an equilibrium. Denote this
common value by $\pi_{eq}$, and the values of $c$ in the two phases
at such an equilibrium by $c_{eq}^\alpha$ and $c_{eq}^\beta$.

By Taylor's expansion about this point of incoherent equilibrium 
(and, in the second line, the use of (\ref{pi'})), 
we have the following approximations 
which are useful in the $\alpha$ phase: 
\bea
\mu(c^\alpha) & = & 
\mu_{eq} + \mu'(c_{eq}^\alpha)(c^\alpha-c_{eq}^\alpha)  +
O(c^\alpha - c_{eq}^\alpha)^2 \lbl{mu-c} \\
\pi(c^\alpha) & = &
\pi_{eq} + c^\alpha_{eq}\mu'(c_{eq}^\alpha)(c^\alpha-c_{eq}^\alpha)  +
O(c^\alpha - c_{eq}^\alpha)^2 \lbl{pi-c}
\eea
where the prime denotes a derivative with respect to $c$.
There is a similar approximation for the $\beta$ phase.
It follows from (\ref{mu-c}) and(\ref{pi-c}) that
\be 
\pi(c^\alpha)
\pi_{eq} + c^\alpha_{eq}(\mu(c^\alpha)-\mu(c_{eq}^\alpha))  +
O(\mu(c^\alpha) - \mu_{eq})^2 
\lbl{pi-mu}
\ee
>From this we can subtract the corresponding formula for $\pi(c^\beta)$, 
and using also the condition (\ref{equalmu}) in the form
$\mu(c^\alpha) = \mu(c^\beta)$, we obtain
\be
[\pi] = [c_{eq}](\mu - \mu_{eq}) + O(\mu - \mu_{eq})^2
\lbl{picmu}
\ee
where $\mu$ denotes the common value of $\mu^\alpha$ and $\mu^\beta$.
Combining (\ref{picmu}) and (\ref{mu-c}), and using (\ref{pikap}), we get
the approximate formula
\be
c^\alpha - c^\alpha_{eq} = 
{\mu - \mu_{eq} \over \mu'(c_{eq}^\alpha)} =
{[\pi] \over [c_{eq}]\mu'(c_{eq}^\alpha)}
\lbl{proto-GT}
\ee
The formula for $c^\beta - c^\beta_{eq}$ is analogous.

For the model alloy we are considering here, the chemical potential
can be expanded in virial-type expansions about the two points 
$c_0^\alpha$ and $c_0^\beta$. If we take $c_0^\beta$ 
to be zero for simplicity, and 
in a case such as Ni$_3$Al 
neglect the small concentration of A atoms on
the sites preferred by B atoms,
these expansions have the form
\bea
{\rm exp}{\mu(c) - \mu_{eq} \over \kk T} & = & 
{c \over c_{eq}^\beta}\{1 + O(c)\}
\nonumber \\
{\rm exp}{\mu_{eq} - \mu(c) \over \kk T } & = & 
{c_0^\alpha - c \over c_0^\alpha - c_{eq}^\alpha}\{1 + O(c_0^\alpha -c)\}
\lbl{muvirial}
\eea
where $\kk$ is Boltzmann's constant and $T$ is the temperature.
The reason why $\mu(c)$ appears with a negative sign in the second equation
is that in the $\alpha$ phase the small parameter of the virial 
expansion is $c_0^\alpha - c$, the concentration
of $B$ atoms, and the relative chemical potential of these atoms is $-\mu(c)$.
Using (\ref{muvirial}) in (\ref{proto-GT}), 
we obtain the approximate formulas
\be
c^\alpha = c_{eq}^\alpha +
{(c_0^\alpha - c_{eq}^\alpha) [\pi] \over 
\kk T[c_{eq}]}, ~~~
c^\beta = c_{eq}^\beta +
{c_{eq}^\beta [\pi] \over 
\kk T[c_{eq}]}
\lbl{GT}
\ee
in which $[\pi]$ stands for the expression on the right side of (\ref{pikap}). 
An example showing how the differential equation and boundary conditions
formulated in this section are used is given in the following section

The formula (\ref{GT}), reducing in the absence of elastic effects
to the Gibbs-Thomson formula, is due to Johnson and Alexander \cite{86joh}
(see also \cite{88rot,89leo1,93abi1,96su1}),
who used it (in a more accurate version) to calculate $c$ at the surface of a
spherical inclusion with isotropic elasticity.

\subsection{Sharp-interface evolution calculations} 

\cs \cs \cs
\subsubsection{Growth or shrinkage of an isolated spherical inclusion}
\lbl{GSISI}

In this section we calculate (following Laraia et al \cite{88lar,89lar})
the rate of growth of a spherical inclusion of radius $R$
in an infinite elastically isotropic matrix, 
with the boundary condition that $c$ approaches
a prescribed limit $c^\infty$ far from the inclusion.  
At the surface of the inclusion its value is given by (\ref{GT}).
The calculations are simple because the geometry makes $[\pi]$, 
and therefore $c^\alpha$ and $c^\beta$, 
independent of position on $\Gamma$, so that they can be treated as constants.

We make the approximation (which appears to be a good one if $R dR/dt \ll D$)
of neglecting the time derivative in the diffusion equation (\ref{diffneq}), 
so that it reduces to Laplace's equation. 
The appropriate solution, with origin at the centre of the sphere, 
is
\be
c =
\left\{
\begin{array}{ll}
c^\alpha                                & (r < R)  \nonumber \\
c^\infty + (c^\beta - c^\infty)R/r      & (r > R)
\end{array} \right .
\ee
where $r$ means $|\bx|$.
Using (\ref{Gammadot}) we find the rate of growth of the inclusion to be
\bea
{dR \over dt} = v_n = -D\left[{dc/cr}\right] 
 =  D(c^\beta - c^\infty){1\over R} 
\lbl{proto-LSW}
\eea                
If there are no elastic forces, then (\ref{pikap}) reduces to
$[\pi] = 2 \sigma/R$, and so (\ref{proto-LSW}) with $c^\beta$ taken from
(\ref{GT}) gives
\be 
{dR \over dt} = {2 \sigma D c_{eq}^\beta \over \kk T[c_{eq}]R}
                \left({1\over R^*} - {1\over R}\right)
\lbl{LSWeq} 
\ee              
where $R^*$, defined by
$2\sigma/R^* = \kk T [c_{eq}](c^\infty - c_{eq}^\beta)/c_{eq}^\beta$ 
is the so-called critical radius, the radius of an inclusion
at whose surface the value of $c^\beta$ would be $c^\infty$.

If elastic forces are present, then eqn (\ref{pikap}) can be evaluated using
the $R_2 \to \infty$ limit of the solution for a spherical inclusion
outlined in section \ref{elementary}. The result is
\be
[\pi] = 2 \sigma/R + 18K^\alpha G^\beta [q]^2/(3K^\alpha + 4G^\beta).
\lbl{[pi]}
\ee 
Eqn (\ref{LSWeq}) still holds, with $R^*$ now defined by
\be
2\sigma/R^* = \kk T [c_{eq}](c^\infty - c_{eq}^\beta)/c_{eq}^\beta
- 18K^\alpha G^\beta [q]^2/(3K^\alpha + 4G^\beta)
\lbl{Rstar}
\ee

Eqn (\ref{LSWeq}) is a central component of the LSW theory of coarsening
at very low volume fractions of precipitate, which we have already 
mentioned and shall discuss further in section \ref{LSW}. 
In that theory each precipitate has its own value of $R$, which changes 
with time according to eqn (\ref{LSWeq}). 
The common value of $R^*$ is determined by requiring the total volume
of the precipitates to be conserved. LSW argue that both $R^*$ 
and the average radius of the precipitates will grow
in proportion to $t^{1/3}$.
Since eqn (\ref{LSWeq}) is unaffected by the presence of the elastic
term in (\ref{[pi]}) it follows that the elastic term has no effect
whatever on the coarsening behaviour as predicted by the LSW theory.
The elastic effects will be revealed only
by going beyond the main assumptions made in the above calculation,
which are that the elasticity is isotropic, the precipitates are
spherical, and their volume fraction is very small.

\subsubsection{A single non-spherical inclusion} \lbl{diff-1}

\cs We have already seen in Section \ref{iso-1} that a spherical 
inclusion, with isotropic elasticity, 
can be thermodynamically unstable if it is softer than the matrix.
If the inclusion is growing, a further instability becomes possible,
corresponding to the one discovered by Mullins and Sekerka \cite{63mul} 
which in the non-elastic case appears as soon as a certain critical radius
(which depends on the rate of growth) is exceeded. 
This type of instability has been investigated using linear theory by 
several authors \cite{89car,89leo2,91onu,93leo}. 

The first application of the sharp-interface model
to nonlinear time evolution of the shape of a precipitate was 
a two-dimensional calculation done by
Voorhees, McFadden and Johnson (1992) \cite{92voo1}, 
using a boundary integral method to solve the diffusion equation
in the matrix (they neglected diffusion inside the precipitates).
\cs They calculated the shape evolution of a single precipitate 
of arbitrary initial shape for an anisotropic
system having the elastic constants of Ni-Al,
choosing the concentration at infinity so that the volume of the
precipitate remained constant and it approached an equilibrium shape.
In contrast to the energy calculations described in Section \ref{aniso-shape}, 
they did not find any symmetry-breaking transitions : the equilibrium shape
always had fourfold symmetry even if the initial shape 
did not, and their precipitates never split into pieces;
but they attribute this to their not having considered such
large precipitates ({\em i.e.} such large values of $R/R_0$) as the other
workers.

\cs Jou {\em et al.} (1997) \cite{97jou} did similar calculations,
but they used isotropic elasticity and chose the concentration at 
infinity so that the precipitate would grow at a prescribed rate. 
The mass influx has a big effect on the shape, 
favouring the formation of dendrites both for relatively hard and
relatively soft precipitates, whereas the equilibrium shapes are 
`squarish'. If there is an external stress field, the dendrites
perpendicular and parallel to the axis of this field grow at 
different rates.

\subsubsection{Two or three precipitates}\lbl{diff-2}

When studying more than one precipitate, or even a single precipitate in
three dimensions, it is a complicated task to follow the detailed
evolution of the boundary surfaces. To avoid the difficulty, some
authors constrain the precipitates to be spherical, determining
the rate of change of the size and position of each sphere from appropriate
integrals of eqn (\ref{Gammadot}) over the surface of that sphere.
Voorhees and Johnson (1988) \cite{88voo}, also Johnson {\em et al.} (1989) 
\cite{89joh}, studied two spherical precipitates in an
isotropic matrix, solving the diffusion equation in the matrix 
using bispherical coordinates. For precipitates that are softer 
than the matrix, they find inverse coarsening if
the ratio of the particle radii is less than about 2.
Diffusion also make the centres of the precipitates move;
this also happens in the non-elastic case, but here the movement
is faster.
Johnson {\em et al.} (1990) \cite{90joh3} 
extend this type of calculation to anisotropic elasticity.
They find inverse coarsening when the
preciptate separation is along a soft direction 
([100] in Ni, [110] and [111] in Mo),
but coarsening rates are enhanced for some other directions of the separation.

Getting away from the assumption of spherical precipitates,
Su and Voorhees (1996a,b) \cite{96su1,96su2}
did 2-D calculations for a pair (or triplet) of inclusions in a matrix
with cubic symmetry, allowing for shape changes
(but not for diffusion inside the precipitates).
Large precipitates adopted approximately rectangular shapes, 
rather than the round ones assumed in fixed-shape calculations. 
Inverse coarsening was never observed for two-precipitate systems
(notwithstanding the results of earlier calculations at fixed shape); 
in systems of three precipitates, local inverse coarsening was observed 
but it was never enough to stabilize the system as a whole
against further coarsening. 
With regard to the movement of the precipitates, the main effect of
the elastic forces was to align pairs of precipitates so that the
line joining their centres was along one of the elastically soft
directions. These forces also caused the precipitates to move
close together (partly by motion of their centres, partly by
shape changes), but not to touch.
In discussing these results the authors make use of the concept of
configurational forces due to Eshelby \cite{51esh} and Gurtin \cite{98gur}

\subsubsection{Many precipitates : the LSW theory}   \lbl{LSW}

\cs 
The first theory of coarsening, which took no account of elastic
interactions, was the LSW (Lifshitz and Slyozov \cite{61lif},
Wagner \cite{61wag}) theory, which is based on eqn (\ref{LSWeq})
for the rate of growth of an isolated spherical precipitate. 
Precipitates that are larger than $R^*$ in radius grow by diffusion
at the expense of those that are smaller.
The theory, which assumes that the volume fraction of precipitates is 
vanishingly small, predicts that the 
average radius of the precipitates present at time $t$ will
grow in proportion to $t^{1/3}$. For finite volume fractions
the LSW theory becomes inaccurate, and it can be improved upon
by following numerically the individual sizes of a system of precipitates
which are assumed to be spherical. 
For reviews of this theory see Voorhees (1985, 1992) \cite{85voo,92voo2}. 

Enomoto and Kawasaki (1988,1989) \cite{88kaw}, \cite{89eno}
extended this type of numerical simulation by including
the effect of elastic forces.
They assumed that the precipitates were spherical,
so that the total free energy could be expressed 
in terms of the radii of all the inclusions, using
Eshelby's formula (\ref{EshelbyE}) for the elastic interaction. 
(If this interaction is not included, then the elastic
energy does not significantly affect the rate of coarsening : 
see \cite{89lar}).
The diffusion potential at the surface of an inclusion was
replaced by an average value over that surface, calculated in  
terms of the partial derivative of the total free energy with
respect to the radius of the inclusion.
For precipitates that were harder ({\em i.e.} more rigid) than the matrix,
they found that the elastic interaction assisted the
coarsening mechanism and indeed speeded it up so that the
average radius eventually grew in proportion to $t^{1/2}$ instead of $t^{1/3}$.
On the other hand, if the precipitates were softer than the  matrix,
the elastic interaction was found to slow down the coarsening or even stop
it altogether. 

A serious disadvantage of the simulations of Enomoto and Kawasaki
is that they depend on Eshelby's formula (\ref{EshelbyE})
for the elastic interaction, which in turn depends on an
assumpton that the precipitates are spherical.
As pointed out by Onuki and Nishimori (1991) \cite{91onu4} 
(see section \ref{iso-energy2} above)
if the precipitates are allowed to change their shape, 
the interaction energy may deviate greatly from this formula 
and so these simulations do not even convincingly establish the
$t^{1/2}$ growth law. 

Shortly after these simulations were done Leo {\em et al.} \cite{90leo}
gave a general scaling argument leading to the conclusion that
when the precipitates are large enough for the surface tension
to be unimportant the average linear size of the precipitates 
will indeed grow in proportion to $t^{1/2}$.
Their argument is based on the same idea as one given by
Mullins and Vi\~{n}als \cite{86mul} \cite{89mul} which
predicts a $t^{1/3}$ growth law
for the case where there are no elastic forces. The
main assumption used in this argument is that the domain structure is
statistically self-similar, {\it i.e.}  that at any time
it has the same statistical
properties as the one obtained by a uniform
expansion of the domain structure at any earlier time.
However, in contrast to the capillarity-dominated case,
where there is plenty of
evidence for the statistical self-similarity hypothesis and the
$t^{1/3}$ law is well-established,
in the case where elastic forces dominate
there appear to be many cases, particularly for anisotropic materials,
where the hypothesis of self-similarity is false.
For example, the hypothesis would require 
the equilibrium shape of a single inclusion to 
be independent of its size, but for anisotropic elasticity this is not so
(see section \ref{aniso-shape} above).
The scaling argument does, however, tend to confirm 
the $t^{1/2}$ growth law found by Enomoto and Kawasaki for spherical
precipitates which were harder than the matrix.

Besides the growth law,
the positional correlations of particles have received
much attention in the non-elastic case. For elastically interacting systems, 
the main studies of positional correlations using the sharp-interface
method have been done by Abinandanan and Johnson (1993). They calculated  
\cite{93abi1,93abi2} the concentration field in a three-dimensional
isotropic matrix containing spherical inclusions, using a multipole
expansion method. (For an alternative to the multipole
expansion method, see ref. \cite{94hor}).
Unlike Enomoto and Kawasaki, they took the elastic stiffness matrix to
be homogeneous, but they avoided the Bitter-Crum theorem
by making the misfit strain tetragonal rather than isotropic.
They obtained the effect of diffusion in the matrix 
(though not inside the precipitates) not only on the sizes 
of the precipitates, but also on their velocities.
The resulting formulas were used in a simulation of the time 
evolution of a system of spherical precipitates whose initial positions
and sizes were chosen from a random distribution. In the final configuration,
the positions of most of the surviving precipitates were strongly
correlated : arranged in a plane perpendicular to the axis of the tetragonal
misfit strain if the axial component of the anisotropic part of the 
misfit strain was an expansion, but in lines parallel to the misfit 
strain axis if this component of the misfit strain was a compression. 
In a later paper  \cite{95abi} Abinandanan and Johnson
discuss the development of spatial correlations during coarsening.
For small particles, capillary forces dominate and the particles
hardly move; the spatial correlation is that each large particle 
is surrounded by a depletion zone from which it has eaten most
of the solute material. For large particles, elastic forces dominate 
and the particles do move, setting up a new system of correlations 
characterized by clustering of favourably oriented similar sized particles.

\hs Hort and Johnson (1996) \cite{96hor2} simulate
the time evolution of a system of
spherical precipitates in the presence
of a uniaxial applied stress, both for homogeneous elasticity with
tetragonal misfit strain and for heterogeneous elasticity with isotropic
misfit strain. The precipitates are found to align themselves
in a way that is consistent with eqn (\ref{raftsign}).

All the calculations mentioned so far in this section depend on the
(dangerous) simplifying assumption of spherical precipitates. 
Jou {\em et al.} (1997) \cite{97jou} do two-dimensional calculations which do
take shape changes into account, using the boundary integral method 
mentioned in section \ref{diff-1}. Their calculations refer to 
an elastically isotropic but heterogeneous material. 
Complicated effects are observed :
alignment, movement, coagulation, coarsening, with everything
depending strongly on the amount of elastic inhomogeneity, the misfit
strain, and the externally applied field.

\section{Diffuse interface models}
\lbl{DImodel}

In this type of model we no longer represent the phase boundaries 
as a geometrical surface. 
Instead, the micro-structure of the interface is itself included in
the model. The concentration $c(\bx,t)$ of the A-component is now treated as
a function of position $\bx$ and time $t$ which is continuous throughout 
the whole of $\Omega$.
The surface $\Gamma$ and the domains $\Omega^\alpha$ and $\Omega^\beta$
play no part in the mathematical formulation of these models, though it
may be possible to identify them approximately once the concentration field
is known. In particular, at late stages of the coarsening process,
large domains develop in which $c$ is almost uniform, being close
to one or other of the equilibrium values $c_{eq}^\alpha$ and $c_{eq}^\beta$
defined in section \ref{SUM}; these domains, which are the analogues
in this model of the domains $\Omega^\alpha$ and $\Omega^\beta$
used in the sharp-interface model, are separated 
by a relatively thin transition layer whose thickness does not 
change with time, and which is the analogue of the geometrical surface
$\Gamma$ in the sharp-interface model.

An advantage of the method is that it gives information about the very early 
stages of spinodal decomposition, before any well-defined interfaces 
have formed \cite{61cah}. For the late stages, it has further advantages:
no a priori assumption about the topology of the
interface need be made -- this topology is determined by the
evolution of the model itself -- and in numerical work the
difficult problem of dealing with moving free boundaries is avoided.
Moreover, even when the topology does not change,
the rather complicated conditions on the surface of $\Gamma$
need not be allowed for explicitly.

\subsection{The Cahn-Hilliard equation and some of its relations}

\subsubsection{The Cahn-Hilliard equation}\lbl{CHEsec}

The diffuse-interface 
method was introduced in Cahn's celebrated paper about spinodal
decomposition (1961) \cite{61cah}. He based his initial discussion on a
formula, due originally to van der Waals \cite{1894vdw},
for the free energy of a non-uniform system without elastic energy
\be
F = \int_\Omega [{f}(c) + \half \chi (\nabla c)^2] d^3 \bx.
\lbl{CahnF1}
\ee
Here $c$ (meaning $c(\bx,t)$) is the concentration field, 
assumed to be a continous (indeed differentiable) function of position, 
${f}(c)$ has the same meaning as in (\ref{free-energy})
and $\chi$ is a positive constant, so that
the gradient term penalizes rapid spatial variations of $c$. 
To describe phase transitions, the function ${f}$ is assumed 
to have a non-convex graph, as illustrated in Fig. 6.
For example the approximation to $f$ given
by the mean-field theory of a binary substitutional alloy,
\be
f(c) = \mu_{eq} c - 2\kk T_0c(1-c) + \kk T \{c \log c + (1-c) \log (1-c)\} 
\lbl{meanfieldf}
\ee
where $\mu_{eq}$ and $T_0$ are constants,
has a non-convex graph when $T < T_0$. 

For the time dependence of $c$, Cahn assumed
\be
{\partial c \over \partial t} = 
\nabla \cdot \left(M \nabla {\delta F \over \delta c(\bx)}\right)
\lbl{Cahnkinetic}
\ee
where $M$ is a mobility coefficient, which may depend on $c$, and
$\delta F / \delta c(\bx)$ is the variational
derivative of $F$ with respect to $c(\bx)$.
The variational derivative is defined by the condition that
\be
\int_{\Omega} {\delta F \over \delta c(\bx)} \phi(\bx) d^3 \bx
= \left({dF( c + \epsilon \phi) \over d \epsilon}\right)_{\epsilon = 0}
\lbl{varderiv}
\ee
for all differentiable functions $\phi$ which vanish on the 
boundary of $\Omega$. In the present case this definition gives
\be
{\delta F \over \delta c(\bx)} = {f}'(c(\bx)) - \chi  \nabla^2 c(\bx)
\lbl{Cahnmu}
\ee
where ${f}'(c)$ denotes the derivative of ${f}(c)$, the same thing as 
the local chemical potential used in section \ref{MD},
and so (\ref{Cahnkinetic}) becomes
\be
{\partial c \over \partial t} =
\nabla \cdot M \nabla ({f}'(c(\bx)) - \chi  \nabla^2 c(\bx))
\lbl{CH}
\ee
This is the Cahn-Hilliard equation, which is the cornerstone of
the theory of spinodal decomposition without elastic interactions.
If the formula (\ref{meanfieldf}) is used for $f$, $M$ is generally
taken to be $Dc(1-c)/\kk T $ where $D$ is the diffusivity (assumed for
simplicity to be independent of $c$ --- compare eqn (\ref{diffneq})), 
so as to make (\ref{CH}) reduce to Fick's law when $c(1-c)T_0/T$ and
$\chi$ are small. 

\cs Eqn (\ref{CH}) is to be solved with boundary conditions
modelling the conditions at the surface of the material, and initial
conditions modelling the initial concentration distribution. 
For example, to represent an alloy quenched from a high 
temperature to a point inside the miscibility
gap of the phase diagram, $c(\bx,0)$ is generally taken 
(see, for example, \cite{89cha}) to be 
$\bar{c} + \zeta(\bx)$, where $\bar{c} = N_A/|\Omega|$ 
is the space-averaged concentration of $A$ atoms and $\zeta$ 
is a spatially random function representing small concentration fluctuations.
If the uniform state $c(\bx) = \bar{c}$ 
is unstable (see section \ref{SUM} below)
these fluctuations grow with time, eventually producing concentration 
distributions very similar to the ones that are observed experimentally
in spinodal decomposition if elastic interactions are unimportant.

For many alloys, further order parameters besides
the concentration are necessary to characterize the phase transition
completely. The prototype of such situations is the order-disorder
transition at constant concentration, as in Cu-Au, where at low
temperatures each kind of atom prefers \cs one of two sub-lattices
of a body-centred cubic lattice.
The mathematical model uses an order parameter field $\eta(\bx)$ which 
is the local concentration of (say) Cu atoms on one of \cs the sublattices.
The free-energy functional has the form (\ref{CahnF1}) with $\eta$
replacing $c$, but since $\eta$ does not obey a conservation law
the kinetic equation is not (\ref{Cahnkinetic}) but
the Allen-Cahn \cite{79all} equation 
\be
{\partial \eta \over \partial t} = -{\Lambda} {\delta F \over \delta \eta(\bx)}
\lbl{AC}
\ee
(also known as the time-dependent Ginzburg-Landau equation)
where ${\Lambda}$ is a kinetic coefficient, which may depend on $\eta$.
In general, both the concentration field and one or more order parameter
fields will appear in the free-energy functional, the concentration
obeying an equation of the form (\ref{CH}) and each non-conserved 
order parameter obeying an equation of the form (\ref{AC})
\footnote{In principle, cross-terms might appear in these equations,
for example, if there are two order parameters, 
$\partial \eta_1 / \partial t = 
- {\Lambda}_{11} (\delta F / \delta \eta_1) + {\Lambda}_{12} (\delta F /
\delta \eta_2))$)} 

A further refinement of these equations, which makes it possible 
to allow for fluctuation effects such as nucleation, 
is to include noise terms on the 
right sides of (\ref{CH}) and (\ref{AC}). 
\cs This development was initiated by Cook \cite{69coo2}.
The resulting equations might be 
called the stochastic Cahn-Hilliard and stochastic Allen-Cahn equations. 
For further information about their use in physics see the reviews by 
Gunton {\em et al.} \cite{83gun1} or Gunton and Droz \cite{83gun2}.
\cs On the mathematical side, it is known that the one-dimensional
stochastic Allen-Cahn equation with space-time white noise
has a solution (indeed a unique solution), 
but in more than one dimension this is known
only for coloured noise. \cite{98gyo}

\subsubsection{Including elastic energy in the Cahn-Hilliard equation} 
\lbl{Onuki}

\cs The free energy formula (\ref{CahnF1}) can be generalized to 
include elastic interactions if we introduce a new field variable
$\bu(\bx)$ and replace $f$ by $f + w$ just as we did in 
(\ref{free-energy}). In order for the elastic interaction to 
have any effect on the diffusion, however, we must allow
$\lambda_{ijmn}$ and/or $e_{ij}^0$ to depend on $c$,
so that the elastic energy density $w$ now depends on $c$ as well as the
strain. The free energy formula thus becomes
\be
F\{c(\cdot),\bu(\cdot)\} = 
\int_\Omega [{f}(c(\bx)) + w(c(\bx),e_{11}(\bx),\dots) 
+ \half \chi (\nabla c(\bx))^2] d^3 \bx
\lbl{CahnF2}
\ee
with $w(c,e_{11},e_{12},\dots,e_{33})$ 
given by the formula (\ref{w}), as before. The $c$-dependence of $w$ 
arises from the possibility
that $\lambda_{ijmn}$ and $e_{ij}^0$ may depend on $c$.
Cahn's recipe (\ref{Cahnkinetic}) now gives, in place of (\ref{CH}), \cs
\bea
{\partial c \over \partial t} & = &
\nabla \cdot M \nabla \left({f}'(c(\bx)) +
{\partial w(c(\bx),e_{11}(\bx),\dots) \over \partial c} 
- \chi\nabla^2 c(\bx) \right)      \nonumber \\
& = & \nabla \cdot M \nabla (\hat{\mu} -\chi \nabla^2 c(\bx))
\lbl{eCH}
\eea
where $\hat{\mu}$ is the {\em diffusion potential}, a function of 
$c$ and $e_{11},e_{12},\dots$ defined by
\bea
{\hat{\mu}} 
& = & {\partial(f(c)+w(c,e_{11},e_{12},\dots)) \over \partial c} \nonumber \\
& = & \mu(c) + \half \sum_{ijmn}{d \lambda_{ijmn} \over d c}
(\Delta e)_{ij} (\Delta e)_{mn}
- \sum_{ijmn}\lambda_{ijmn} (\Delta e)_{ij}
{d e^0_{mn} \over d c}
\lbl{hmu}
\eea
Equation (\ref{eCH}), without the gradient term, is due to
Larch\'{e} and Cahn \cite{82lar}. The complete equation 
was given by Onuki (1989) \cite{89onu1}. 

\cs In using eqn (\ref{eCH}) the elastic field variable $\bu$ is assumed 
to relax to its equilibrium value much faster than the concentration, 
and therefore it is determined, just as in the sharp-interface model, 
by minimizing $F$. The Euler-Lagrange equation
for the minimization is, as in the sharp interface
model, eqn (\ref{eleqm}). 
In consequence of this minimization, the strain at each point 
in the material is a (non-local) functional of 
the entire density profile $c(\cdot)$ and so the 
diffusion potential at that point is likewise such a functional.
For an example, see eqn (\ref{muexample}) below.

As in the sharp interface model, the minimization to find $\bu$ is easiest  
when the elastic stiffness tensor is independent of $c$ and therefore
uniform, so that the Fourier transform formula (\ref{LRint}) can be used.
A convenient way to allow for the dependence of stress-free strain on
concentration in this formula 
is to assume a linear relation analogous to Vegard's law :
\be
e_{ij}^0 = a_{ij} + b_{ij} c
\lbl{preVegard}
\ee
where $a_{ij}$ and $b_{ij}$ are consants which can be expressed in 
terms of the quantities $e_{ij}^\alpha$, $c_{eq}^\alpha$ etc. 
used in chapter 2.
By substituting this into eqn (\ref{LRint}) and using the result in
(\ref{CahnF2}) we can express the free energy as a functional of the
field $c$ alone :
\be
F\{c(\cdot)\} = 
\int_\Omega \{f(c(\bx)) + \half \chi (\nabla c(\bx))^2 \} \dd^3 \bx
+ \half \int \int_\Omega 
(c(\bx) - \bar{c}) V_{el}(\bx - \by) (c(\by) - \bar{c}) \dd^3 \bx \dd^3 \by
\lbl{CahnF3}
\ee
where $\bar{c}$ is the space average of $c$, and $V_{el}$ is 
defined by
\be
V_{el}(\bz) = \sum_{ijmn} b_{ij} H_{ijmn}(\bz) b_{mn}
\lbl{Vel}
\ee
Apart from a numerical factor $[c_{eq}]^{-2}$, $V_{el}$ is the inverse Fourier
transform of the function $B$ in Khachaturyan's formula (\ref{khachB}).
The diffusion potential at $\bx$ when $F$ is given by (\ref{CahnF3}) 
is the functional derivative of the non-gradient part of $F$ with respect to
$c(\bx)$ , i.e.
\be
\hat{\mu}(\bx) = 
\mu(c(\bx)) + \int_\Omega V_{el}(\bx-\by)(c(\by)-\bar{c})\dd^3 \by
\lbl{muexample}
\ee 
wwhere $\mu(c) = f'(c) $ is the chemical potential as in (\ref{sfmu})

To obtain information about coarsening from such models, one must
integrate (\ref{eCH}) 
over long time intervals. Methods of integrating the non-elastic
version of this equation are reviewed in \cite{89ell}.
The elastic calculation is a formidable one, particularly in 
three dimensions, since to find $\partial w/\partial c$ the equation
of elastic equilibrium (\ref{eleqm}) must be solved at each time step
--- either as a partial differential equation in position space,
or (if the elastic constants are independent of concentration)
by the Fourier transform formula (\ref{khachw}) or an equivalent
position-space formula such as the double integral in (\ref{CahnF3}). 
Until recently it was not possible to calculate time evolution from 
these equations with any accuracy in a reasonable time, 
and inaccurate or uncertain methods had to be used instead
\cite{88oon,89gay}; but more reliable methods are now becoming available
\cite{98che,98leo1}.

\cs In the limit where $\chi \to 0$, with a suitable concurrent
re-scaling of the time variable, the Cahn-Hilliard equation 
(\ref{CH}) converges to the
non-elastic version of the diffusion problem with moving
interfaces, formulated in chapter \ref{MD}. This was shown formally by
Pego \cite{89peg} and rigorously by Alikakos et al \cite{94ali}.
The corresponding convergence problem for the elastic version, eqn 
(\ref{eCH}), is treated by Leo {\em et al.} (1998) \cite{98leo1}

\subsubsection{Lattice differential equations} \lbl{LDE}

\ds The Cahn-Hilliard equation does not take explicit account of the
lattice structure of the crystal. \cs In the case where the elastic
stiffness tensor is uniform, it is possible to take account
of this lattice structure, without departing from the general
philosophy of the method, by replacing the integral in
\cs (\ref{CahnF3}) by a sum over \cs $N$ lattice sites situated
at postions $\br$, so that the 
free energy $F$ is no longer a functional but a function of 
$N$ variables $c(\br)$: 
\be
F = {|\Omega| \over N}\{ \sum_{\br} f(c(\br)) 
+ \half \sum_{\br} \sum_{\br'}V(\br-\br')(c(\br)-\bar{c})(c(\br')-\bar{c})\}
\lbl{discreteF}
\ee
where $V$ is defined by
\be
V(\bz) = \Delta(\bz) + V_{el}(\bz)
\lbl{V}
\ee
with $\Delta(\bz)$ chosen so that $\sum_{\br'} \Delta(\br-\br')c(\br')$
is a finite-difference approximation to \cs $\chi \nabla^2 c(\br)$.
For example, on a simple cubic lattice with lattice spacing $a$, 
we could take $\Delta(\br) = \chi/a^2$ for the 6 lattice points 
nearest to the point $\br={\bf 0}$, $\Delta({\bf 0}) = -6\chi/a^2$, 
and $\Delta(\br) = 0$ for all other lattice points.
However, once the free energy has been written in the discrete
form (\ref{discreteF}), new possibilities become available, since
the non-elastic forces can now be allowed for explicitly instead
of through their contribution to the thermodynamic free energy
density $f(c)$. For example, $V(\bz)$ could be chosen to be negative 
(favouring unlike pairs of atoms) if the vector $\bz$ connects 
a nearest-neighbour pair of sites but positive 
(favouring like pairs) if $\bz$ connects a next-nearest neighbour pair.
\ds Free-energy formulas of this type are the basis of
Khachaturyan's discussion \cite{62kha,63kha1,63kha2} of equilibrium 
structures in crystals. 

The discrete version of the Cahn-Hilliard kinetic equation 
(\ref{Cahnkinetic}) is 
\be
{\partial c(\br,t) \over \partial t} =
\sum_{\br'} M(\br - \br') 
\left({\partial F \over \partial c(\br')} 
    - {\partial F \over \partial c(\br)}\right)
\lbl{Khacheq}
\ee
where $F$ is given by (\ref{discreteF}) and 
the translationally invariant `mobility matrix' $M(\br - \br')$ 
takes a value proportional to the mobility $M$ used in (\ref{Cahnkinetic}). 
Eqn (\ref{Khacheq}) is due to Khachaturyan \cite{67kha,83kha}
and, independently, \cs Hillert, Cook {\em et al.} \cite{61hil,69coo}. 
It was introduced into this field by Chen and Khachaturyan 
\cite{67kha,91che1,91che2}. 

A difficulty with most derivations of (\ref{Khacheq}), 
including the one just given, is that
it is not obvious how $M(\br-\br')$ should depend on $c(\br)$ and $c(\br')$
(corresponding to the factor $c(1-c)$ in the continuous-space formula
$M=Dc(1-c)/\kk T$ given just after eqn (\ref{CH})). 
There is a microscopic derivation (Binder (1974) \cite{74bin},
Penrose (1991) \cite{91pen}) leading to a more complicated kinetic equation
which, if $c(\br)$ does not vary too rapidly from one
lattice site to the next, reduces to (\ref{Khacheq}) with $M(\br-\br')$
approximatetely proportional to $Dc_{av}(1-c_{av})/\kk T$ 
where $c_{av}$ is the local average $\half(c(\br) + c(\br'))$.
In actual computations using (\ref{Khacheq}) with the elastic term
$M(\br-\br')$ has, however, been given the value $D\bar{c}(1-\bar{c})/\kk T$,
independent of the local value of $c$.
The results of these computations are discussed in section \ref{SDI} below.

\subsection{The stability of a uniform mixture}\lbl{SUM}

\subsubsection{Isotropic elasticity} \lbl{iso-stab}

It was pointed out by Cahn (1961) \cite{61cah}, that elastic interactions
tend to stabilize a homogeneous mixture which on purely thermodynamic
grounds might be expected to separate into two phases. In his original
paper he showed that this mechanism would work even in an isotropic solid
\footnote{The `Bitter-Crum' theorem
might seem to ilmply that isotropic homogeneous elasticity
never has any effect, but this is not so: 
what it really tells us is that in a mixture
of two phases $F$ depends only on the overall mole fraction, not
on the geometry \cite{84cah}, a fact which is perfectly consistent
with eqn (\ref{iso-stab1}). 
}.

The mixture is assumed to obey Vegard's law, that the stress-free
strain is isotropic and varies linearly with the concentration, so that
\be
e_{ij}^0 = \eta(c-c_0)\delta_{ij}
\lbl{Vegard}
\ee
where $\eta$ and $c_0$ are constants. Then we can insert 
the formula (\ref{iso-khachw})
for the total elastic energy into (\ref{CahnF2}) and obtain,
in the absence of externally applied strain,
\be
F = \int_\Omega (f(c) +
{2E \over 1-\nu} (c(\bx)-\bar{c})^2 + \half \chi (\nabla c)^2) \ddx
\lbl{iso-stab1}
\ee
where $\bar{c} = \Omega^{-1}\int_\Omega c(\bx) \dd^3 \bx$.
The uniform state with $c(\bx) = \bar{c}$ everywhere is stable if $F$ is
a minimum with respect to variations of $c$
satisfying the constraint  $\int (c-\bar{c}) \dd^3 \bx = 0$.
A sufficient condition for such a minimum is for the
function $f(c) + (2E/(1-\nu)) (c-\bar{c})^2$ to be locally
convex near $c=\bar{c}$,
and if $2E/(1-\nu)$
is large enough this can happen even though $f(c)$ is not locally convex,
so that without the elastic interaction the mixture would be unstable.
The uniform state is stable against all such variations if and only if
the function $f(c) + (2E/(1-\nu)) (c-\bar{c})^2$
is convex, and it is stable against small variations if and only if
this function is locally convex near $c=\bar{c}$, that is to say if
\be
f''(\bar{c}) + 2E/(1-\nu) > 0
\lbl{iso-stab2}
\ee
where $f''$ denotes the second derivative of the function $f$.
The elastic term can make the left-hand side of (\ref{iso-stab2}) positive
even though $f''(\bar{c})$ is negative, and so
the coherent mixture can be stable against phase separation
even though an incoherent mixture of the same composition would not be.
This effect can lower the critical temperature of such mixtures significantly.
\footnote{Khachaturyan, on page 490 of reference \cite{83kha} contests this
conclusion and gives a formula implying that for an isotropic
crystal the stabilizing effect is zero. The difference of opinion
appears to arise from different definitions of \cs the elastic energy: 
according to our definitions the elastic energy of an alloy without 
externally applied stress is zero if the concentration is uniform,
since the mininum of $\int w \ddx$ is zero, achieved by 
taking $e_{ij} = e_{ij}^0$ everywhere (see eqn (\ref{w})); 
whereas according to Khachaturyan's definition part of the 
energy is elastic energy even when the concentration is uniform.}
\cs For a discussion of this and related effects using 
a simple thermodynamic model see Cahn and Larch\'{e} (1984) \cite{84cah}

\subsubsection{Anisotropic elasticity} \lbl{aniso-stab}

The above treatment of stability is easily generalized to anisotropic
elasticity 
(Cahn (1962) \cs \cs
\cite{62cah1}).
Substituting Vegard's law (\ref{Vegard}) into (\ref{khachw})
we obtain, in the absence of externally applied stress,
\bea
W & = & {1 \over 2|\Omega|}
\sum_{\bk \ne 0} \sum_{im} \eta^2 |c(\bk)|^2 \Psi_{iimm}(\bk) \nonumber \\
& \ge & {\eta^2 \over 2|\Omega|} \sum_{\bk \ne 0} |c(\bk)|^2
\min_\bk  \sum_{im} \Psi_{iimm}(\bk) \nonumber \\
& = & \half \eta^2  \min_\bk  \sum_{im} \Psi_{iimm}(\bk)
\int_\Omega (c(\bx) - \bar{c})^2 \dd^3 \bx ~~~\hbox{by~Parseval's~theorem}
\lbl{aniso-stab1}
\eea
As in the discussion of (\ref{iso-stab1}), it follows that the uniform state
is stable against small concentration fluctuations if
\be
f''(\bar{c}) + \half \eta^2  \min_\bk  \sum_{im} \Psi_{iimm}(\bk) > 0
\lbl{aniso-stab2}
\ee
\cs This equation, in the form appropriate to a cubic crystal, is 
due to Cahn \cite{62cah1}. 
The elastic term is always positive (see the text just after eqn (\ref{sG}) 
above) and therefore the elastic interactions tend to stabilize the mixture
against phase separation, just as they do for isotropic elasticity.

A generalization of this kind of stability analysis, with the elastic
moduli depending on concentration, has been used by Thompson and Voorhees
\cite{97tho} to discuss the effect of an externally applied stress
on orientation of the newly-formed precipitates in an anisotropic alloy.

\subsection{Simulations with the diffuse interface model}\lbl{SDI}

\subsubsection{A single precipitate}\lbl{DI-1}

Wang {\em et al.} \cite{91wan1} carried out two-dimensional 
calculations for a single precipitate in a cubic solid 
with negative isotropy, using Khachaturyan's kinetic equation
(\ref{Khacheq}). The initial configuration was chosen to 
represent a single circular precipitate, and they waited
for an equilibrium to be reached, for various different
values of a parameter representing the ratio of the 
elastic to the short-range interactions in their model.
Allowing for the difference between two and three dimensions,
this parameter is proportional to the one denoted by $1/R_0$
in section \ref{scope} of the present paper.
As the parameter was increased (which corresponds to considering
larger and larger precipitates in a real physical system) 
the equilibrium shape,
at first round, became more and more square; then, a hole
appeared in the middle of the square and finally it split into
two rectangles. It is also possible for the faces of the
`square' to become concave \cite{95wan1}

In three dimensions there are even more possibilities, particularly
if the transformation strain is anisotropic 
(Wang and Kachaturyan 1996 \cite{96wan}).
\ds However, as these authors show in \cite{95wan4} 
the behaviour of isolated precipitates is not necessarily 
a reliable guide to their behaviour in the presence of others.

\subsubsection{Many precipitates} \lbl{DI-many}

Nishimori and Onuki (1990) \cite{90nis} \cs\cs
carried out two-dimensional simulations using the elastic Cahn-Hilliard 
equation (\ref{eCH}) with an anisotropic stiffness tensor.
Starting from an initial concentration distribution corresponding to a 
homogeneous disordered state (\cs {\em i.e.} a constant plus a small 
spatially random term), they found that the shapes of the precipitates
were strongly affected by the elastic anisotropy,
and that they tended to align themselves along the crystal axes. 
See Fig. 7 d,e. 

\cs Wang {\em et al.} \cite{91wan2,92wan1,93wan} \ds (see also \cite{97wan}), 
also did simulations on this type of system, but they used
the Khachaturyan equation (\ref{Khacheq}) and paid more attention
to the effect of volume fraction and to to the visible details of 
the evolution process.
For small volume fractions,
the latest observed 
state consisted mostly of nearly round precipitates near the sites of a
a square superlattice. 
This configuration was reached by a form of inverse coarsening,
in which the precipitates that were close to sites of the nascent
superlattice grew at the expense of those that were not,
regardless of the initial sizes of these precipitates. 
For larger volume fractions (say 50 \%) the
precipitates, initially disconnected, tended to coalesce, 
and the latest observed 
state consisted of needle-like precipitates parallel to the
two soft directions. 
In a later paper \cite{95wan3}, they consider 
a case where the transformation strain is tetragonal rather than isotropic 
(\ds modelling Mg-stabilized cubic ZrO$_2$):
intitially, the precipitates line up along crystallographic
directions; later, an alternating band structure develops \ds \ds 
as the precipitates in some rows grow at the expense of those in others.

\ds Onuki and Nishimori (1991,1992) \cite{91onu3,92nis} 
were the first to investigate
the effects of elastic inhomogeneity by means of this type of simulation.
\cs To isolate the effect of the inhomogeneity (which, confusingly, they
call `elastic misfit' in some of their papers) 
they assumed isotropic elasticity.
If the softer (less rigid) phase occupied less than half of the total
volume, then at early times the configuration
consisted of isolated precipitates of softer phase embedded in a matrix
of the harder phase, but as coarsening proceeded these precipitates 
joined up to make a percolated network structure, so that it was now
the harder phase which was arranged in isolated inclusions, while the
softer phase formed the surrounding matrix (see Fig. 7(a-c)).
Moreover, the growth of these precipitates was anomalously 
slow, {\em i.e.} slower than the usual $t^{1/3}$ law.
Later, these authors did simulations of the same type 
for anisotropic (but still inhomogeneous) elasticity \cite{91nis}.
The anisotropy made the inclusions align themselves parallel to one
of the crystal axes, but still the softer phase wrapped the harder one.
If an external stress in an oblique direction was applied, 
the preferred directions of the inclusions tilted so as to be more
nearly perpendicular to one of the principal axes of the external stress.

The numerical method used by Onuki and Nishimori for the time evolution
was not an accurate one; later work using accurate methods 
(Leo {\em et al.} 1998 \cite{98leo1}) has, however, confirmed their 
conclusions about the reversal of the topological 
roles of the two phases.

These topological changes, which the sharp-interface model 
had completely failed to predict, can be understood qualitatively 
on the basis of the formula for spherical inclusions given in 
section (\ref{elementary}), according to which the energy
per unit volume of precipitate is less for rigid precipitates
in a flexible matrix than for flexible precipitates in a rigid matrix.
Or, even more qualitatively, consider the extreme case where
both phases have the same bulk modulus but one of the phases
is perfectly flexible. i. e. its shear modulus is zero. If the
inclusions are perfectly flexible the rigidity of the surrounding
matrix will prevent them from
taking up the sizes where their density has its stress-free value
and therefore they will have some strain energy; but
if the matrix is perfectly
flexible then the (rigid) inclusions can take up their stress-free sizes
and shapes and the matrix can expand or contract to its stress-free
density, so that there is no strain energy at all. Thus the energy is lower
for a perfectly flexible matrix than it is for perfecty flexible inclusions.

\subsubsection{Ordered phases}

In alloys such as Ni-Al one of the phases (in this case the Ni$_3$-Al
phase) can have more than one variant. Such systems can be treated
using Khachaturyan's kinetic equation (\ref{Khacheq}), generalized
by incorporating additional order parameters as described at the end of
section \ref{CHEsec}. Simulations of various types of alloy requiring this
type of treatment have been done by \cs Sagui et al (1994) \cite{94sag},
by Wang and Khachaturyan (1995) \cite{95wan4} ,
by Li and Chen (1998) \cite{98li1,98li2}
and \ds by Wang {\em et al.} (1998) \cite{98wan}. 
An important new possibility is that the transformation strain can
be anisotropic and hence different in the different variants; 
then, when an external stress is applied, one variant will grow
at the expense of another. In this way, since the conditions of
Eshelby's theorem (see section \ref{theorems}) do not apply, rafting can
take place even with homogeneous elasticity.

A recent paper of this type by Le Bouar {\em et al.}
\cite{98bou} models the time evolution of  
alloys in which the matrix is a cubic crystal but the precipitates
are tetragonal, using a generalized stochastic Cahn-Hilliard equation
with three dependent variables : the concentration and two
order parameters. The theoretical model is spectacularly successful : 
see Fig. 2.

\subsubsection{Thin films}


In this section we look at the possibility of elastic interactions
arising from a violation of condition 4 of the Bitter-Crum theorem,
that is to say in a system with boundaries. An effect of this
type was noted by  Larch\'{e} and Cahn \cite{92lar1}
who pointed out that if spinodal decomposition takes place in
in an isotropic crystal with uniform elastic constants
the equilibrium state can have the two phases on opposite
faces of the film, and because of the elastic misfit the film will bend.
\cs  This effect has indeed been observed, in the $\alpha-\alpha'$
phase separation of Nb-H \cite{79zab,80zab}. In these experiments,
the hydrogen-rich $\alpha'$ phase was found on the outer side
of simply curved specimen foils. In some cases the foil was 
even observed to split into two pieces with half thickness.
Both halves were strongly curved. These results were interpreted
as a spinodal decomposition process where the specimen thickness
was equal to half the spinodal wavelength for the simply curved foil
and exacctly to this wavelength for the split foil

Putting an appropriate approximation for the elastic strain field into 
the elastic Cahn-Hilliard 
equation (\ref{eCH}) Larch\'{e} and Cahn obtained
the kinetic equation
\be
{\partial c \over \partial t} =
\nabla^2 \left[{df(c) \over dc}
+ {2 \eta^2 E \over 1 - \nu}
\left(c-\bar{c} - {3x \over 2h} M_1^c \right) - \chi \nabla^2 c \right]
\lbl{mesoLC}
\ee                                      
where $\eta$ is the compositional expansion coefficient, 
defined in (\ref{Vegard}), $E$ and $\nu$ are Young's modulus
and Poisson's ratio, $\bar{c}$ is the average concentration,
and $M_1^c$ is the first moment of the concentration distribution in the
direction at right angles to the film, defined by
\be
M_1^c = {1 \over h^2} \int_{-h}^{h} (c-\bar{c})x \dd x
\ee
Here $h$ is half the thickness of the film and the coordinate system in the
material is chosen to make the two faces of the unstrained film
the planes $x=h$ and $x=-h$. The one-dimensional version of these
equations, obtained by neglecting the dependence of $c$ on $y$ and $z$,
was studied both 
analytically and numerically by Larch\'{e} and Cahn, and numerically
in more detail by Cahn and Kobayashi \cite{95cah}. The most striking
result was that the elastic term accelerated the separation into two phases.

Another bounded-system problem to which the
diffuse-interface model can be applied
is the growth of thin films by deposition, an
important process in the semiconductor industry. The elastic problem is
more complicated than before because 
there is now a misfit between the film and
the substrate as well as the one due to compositional nonuniformity. 
Moreover, the possibility of surface diffusion \cite{57mul,89sro}
and the effect of material
arriving at the surface have to be allowed for.
At the time of writing the implications of linear stability theory
are still being worked out \cite{91spe,93spe1,95guy,96guy,98guy,98leo3},
though nonlinear effects are clearly important too \cite{93spe2}.

\section{Atomic lattice models}\lbl{ALM} \ds

\subsection{The Born model} \lbl{Born}

In this approach continuum elastic theory is abandoned;
instead the elastic properties of the material
are obtained by considering the mechanical forces between 
individual atoms. Each atom is assiged a definite site $\bp$
on the lattice, whose position when the lattice is 
undistorted we denote by $\bx_\bp$, and 
the displacements $\bu_\bp$ of the atoms from these sites are assumed
to be small enough to justify using the harmonic approximation, 
{\em i.e.} treating the force on each atom as a linear function of 
all the atomic displacements of all the atoms.
The way to calculate the macroscopic elastic 
properties of a crystal from the interatomic
force parameters is explained in the book by Born and Huang (1954) 
\cite{54bor}, and, by doing the calculation in reverse, the interatomic
force parameters can be obtained from phonon dispersion curves. 
For simplicity, we assume that only two-body forces 
act, so that atoms which are close enough together to interact
can be thought of as being joined by springs.
However, one should not be misled by this picture into assuming
that all the atomic interactions are necessarily central forces;
this assumption would imply certain relations between the elastic 
constants (sometimes called Cauchy relations) \cite{54bor} which are violated
in many real materials. For example in a crystal with cubic symmetry the
Cauchy relation is $C_{12} = C_{44}$ which, although a good approximation
for \ds \ds some ionic crystals, is a very bad one in most metals.

In the case of an alloy, the interaction between two atoms depends
not only on their separation but also on their nature. 
It is usual to think of their interaction energy as the sum of 
two parts, one of which, the so-called chemical interaction,
depends only on what kind of atom they are and not on
their positions; the other part, the elastic interaction,
does depend on their positions and has a minimum value of
zero which is achieved for some relative position.
This energy-minimizing relative position depends on the nature of 
the two atoms : one can think of the two kinds of atoms
as being of different sizes, and the natural length of the
spring joining the centres of two atoms the sum of the
radii of the atoms. The stiffness of the spring
may also depend on the nature of the two atoms.
The theory of the equilibrium properties of solid solutions based on 
this type of model has been developed by Khachaturyan \cite{67kha,83kha} 
and by Cook and de Fontaine (1969,1971,1972) \cite{69coo1,71coo,72def}. 
For example Seitz and de Fontaine (1978) \cite{78sei} 
have used the model to do energy calculations for various arrangements
of \cs a pair of precipitates in Al-Cu alloys; the results indicated a
tendency for  
the precipitates to stack as parallel disks, providing a possible 
explanation of the oberved behaviour of supersaturated Al-Cu solid solutions.

Diffusion can be represented in this model by allowing atoms of
the two different kinds to exchange places from time to time,
though sufficiently rarely to allow the lattice to come to equilibrium
under the new forces \ds \ds after each exchange.
These exchanges can be simulated by the Monte Carlo method known
as Kawasaki dynamics, in which the following sequence of steps is
carried out a large number of times : (i) choose a pair of neighbouring 
atoms at random, and also a random number $\xi$ uniformly distributed 
between 0 and 1,
(ii) calculate the increase in free energy $\Delta F$ that would occur if
the two atoms were exchanged, and (iii) carry out the exchange if
and only if $\xi < p(\Delta F)$, where the transition probability
function $p(\cdot)$ satisfies the detailed balance condition
$p(x)/p(-x) = \exp(-x/\kk T)$. 
\cs The change in free energy consists of two parts : one is the change in
(mean) energy which is equal to the change in the minimum potential
energy as a function of the atomic displacements, since the vibrational
energy is an unchanging $\kk T$ per degree of freedom. The other part is
an entropy term depending on the vibrational spectrum of the lattice,
which is generally neglected. Indeed, this term is zero
if the atomic stiffness parameters are independent
of the way the two kinds of atoms are arranged on the lattice sites,
a situation which makes the material elastically homogeneous.
Just what part the entropy term should play in models 
capable of representing heterogeneous elasticity is unclear.
\cs \cs


We shall describe two types of simulation in which Kawasaki dynamics
has been used to follow the diffusive motion of the atoms; 
one using central forces and heterogeneous elasticity,
and the other using noncentral forces and homogeneous elasticity.


\subsection{Central forces}

\subsubsection{The model}\lbl{cfmod}

In two dimensions, the elastic stiffness tensor of any crystal 
with hexagonal symmetry is 
isotropic; moreover, if only central forces act, then there is
a Cauchy relation connecting the elastic constants which is equivalent
to $\nu $ ({\em i.e.} Poisson's ratio) $= 1/4$ \cite{74hoo}. 
In three dimensions, the elastic stiffness tensor of 
\ds a harmonic crystal with fcc symmetry and only central forces is likewise
isotropic \cite{97lee} and again Poisson's ratio is 1/4. 
Lee \cite{95lee,97lee}
took advantage of these circumstances to simulate inclusions
in a isotropic matrix using an atomic lattice model. 
He took the elastic energy to be the sum of the energies of
a set of longitudinal harmonic springs, one for each pair of nearest neighbour
sites, and both the stiffness and the natural length of each spring
depended on the nature of the atoms at its two ends. 
In symbols, \ds \ds the elastic energy is
\be
W = {1 \over 4} \sum_\bp \sum_{\bp'} 
{L}(\bp-\bp', \gamma_\bp,\gamma_{\bp'})
(r_{\bp,\bp'} - l(\bp-\bp',\gamma_\bp,\gamma_{\bp'}))^2 
\lbl{cfW}
\ee
where sums go over all the sites on some lattice, 
 $\gamma_\bp$ is defined to to be $+ 1$ if 
site $\bp$ is occupied by an $A$ atom 
and $-1$ if it is occupied by a $B$ atom,
${L}(\bp-\bp',\gamma,\gamma')$ and $l(\bp-\bp',\gamma,\gamma_)$ 
are the stiffness and natural length of the spring connecting 
an atom of type $\gamma$ at site $\bp$ to an atom of type $\gamma'$
at site $\bp'$, and
$r_{\bp,\bp'} = |\bx_\bp -\bx_{\bp'}| + 
(u_{\bp'} - u_{\bp})\cdot(\bx_\bp - \bx_{\bp'})$ 
is the linearized distance between the atoms at sites $\bp$ and $\bp'$.

\subsubsection{The results} \lbl{cfres}

For a system of two inclusions in two dimensions, Lee found \cite{95lee}
that if the inclusions were softer than the matrix they deformed
and eventually coalesced; if they were harder, they remained circular
and the larger one grew while the smaller one shrank to zero.
This work was extended to anisotropic elasticity in ref \cite{96lee},
with similar results. 

In ref \cite{97lee} Lee (1997) studies
the effect of elastic inhomogeneity on a system of two inclusions for which the
misfit strain and the elastic constants can both be different.
If one inclusion is hard ({\em i.e.} harder than the matrix)
and has a misfit strain while the other is soft one and has no misfit,
or a misfit of the opposite sign, then at equilibrium 
(in two dimensions) the soft inclusion surrounds the hard one. 
But if the soft inclusion has a misfit strain and the hard inclusion
has none, the inclusions move away from one another; and if the two
inclusions have misfit strains of the same sign, the soft inclusion
partially wets the hard one but the rest of the soft inclusion tries to get
away. Similar results are obtained in three dimensions.
If an external stress is applied then the
inclusions change their shape in a way compatible with 
eqn (\ref{raftsign}). 
Reviews of the work described in this section are given in 
\cite{97lee2} and \cite{98lee}.

\subsection{Non-central forces}\lbl{NCF}




\subsubsection{The model}\lbl{ncfmod}

Fratzl and Penrose (1994) \cite{95fra1}  \cs \cs describe a 
model for use in atomic lattice simulations 
which is intended to represent coarsening in a 
cubic crystal with realistic elastic constants. 
For reasons explained earlier in this chapter, 
it is essential to include some non-central forces in the model.
\ds \ds This can done by adding to the right side of the `longitudinal'
elastic energy formula (\ref{cfW}) a term representing transverse springs
\be
{1 \over 4} \sum_\bp \sum_{\bp'} T(\bp-\bp')
\{(\bu_\bp -\bu_{\bp'}) \wedge (\bx_\bp - \bx_{\bp'})/|\bx_\bp -
\bx_{\bp'}|\}^2 
\lbl{transW}
\ee
In these simulations, a square lattice was used, 
with elastic forces which can be pictured by imagining
the atoms connected by three different types of spring :
longitudinal springs connecting nearest neighbors, 
transverse springs connecting nearest neighbors,  
and longitudinal springs connecting 
next-nearest neighbours. To model the elastic misfit, the
natural length of a longitudinal spring connecting sites $\bp$ and $\bp'$
was made to depend linearly on $\gamma_\bp + \gamma_{\bp'}$,
but the stiffnesses of the springs were taken to be
independent of what type of atoms they connected.
This model is characterized by three stiffness
constants, the same as the number of elastic constants in a cubic
crystal. It is possible, in fact, to relate the microscopic stiffnesses
to the macroscopic elastic constants. Formulae for different
lattices are given in, for example, \cite{69coo1,83kha,95fra1}.

\ds \ds Given any configuration $\gamma(\bp)$ of $A$ and $B$ atoms on the
lattice, the minimum of the elastic energy with respect to the atomic
displacements $\bu$ can be found as a function of the $\gamma(\bp)$'s.
The elastic free energy differs from this minimum by
a constant, which can be ignored. 
Since the elastic constants are taken to be homogeneous,
this calculation can be done using Fourier transforms \cite{69coo1,83kha},
in a similar way to the derivation of Khachaturyan's formula (\ref{khachB})
given in section \ref{FTS} (eqns (\ref{FTs})-(\ref{LRint})). The result 
can be written
\be
W = \frac{1}{2N} \sum_{{\bf k}\neq 0} \sB({\bf k}) \;|\tilde{\gamma}({\bf
k})|^2 + {\rm const.}
\lbl{Eel} 
\ee
where the sum extends over the first Brillouin zone of the lattice and
$\tilde{\gamma}$ is the Fourier transform of $\gamma(\bp)$ :
\be
\tilde{\gamma}({\bf k}) = \sum_{{\bf p}} \gamma({\bf p}) e^{i\,{\bf p}
\cdot {\bf k} }
\lbl{Sdef}
\ee
\ds The `elastic potential' $\sB({\bf k})$ can be
calculated for any type of lattice, and has been given explicitly for some,
e.g. in \cite{69coo1,95fra1}.

The elastic energy (\ref{Eel}) can be written in an alternative form,
as a sum of interactions between pairs of lattice sites
analogous to the double sum in eqn (\ref{discreteF}).
Adding to it the chemical energy, the important part of which can also be 
expressed as such a sum, we obtain for the total free energy
\be
F = W_0 + \frac{1}{2} \sum_{{\bf p}}\sum_{{\bf p}'}
(V_{el}({\bf p}-{\bf p}') + V_{chem}({\bf p}-{\bf p}'))
\, \gamma({\bf p}) \, \gamma({\bf p}')
\lbl{FPF}
\ee
where $W_0$ is a constant, 
$V_{el}({\bp})$ is the inverse Fourier transform of $\sB({\bf k})$,
(in which we take $\sB({\bf 0}) = 0$, so that $\sum_\bp V_{el}(\bp) = 0$)
and $V_{chem}(\bp - \bp')$ represents the short-range 
chemical interaction between two like atoms on sites $\bp$ and $\bp'$.


The effects of external stress can also be treated to some extent
within this framework \cite{95lab,97lab}. The treatment requires, however, 
a modification of the model 
since according to Eshelby's theorem (see section \ref{theorems})
no effect will be seen for homogeneous elasticity.
This modification is to assume that the stiffnesses of the springs 
depend weakly on what kind of atom they connect, and to treat
the nonuniformity as a perturbation --- for details see \cite{97lab}. 
In analogy with the analysis for an elastic medium summarized at the end of 
section \ref{FTS}, the elastic energy can be expressed as the sum of three 
parts, one of which is the non-quadratic `Eshelby' interaction.
To keep the problem mathematically simple, this non-quadratic interaction 
was neglected in \cite{97lab}. 

\subsubsection{The results} \lbl{ncfres}

The model just described can be used for simulations in various ways.

When the chemical interaction is taken to be a nearest-neighbor 
attraction 
and the elasticity has cubic anisotropy, stripe-like
domains develop under the influence of the lattice misfit \cite{96fra}. 
The stronger the misfit or the stronger the anisotropy of the elastic
constants, 
the faster the anisotropy of the precipitates increases.
The mean domain size, however, \cs (defined as the ratio of 
area to perimeter) always increases in proportion to
$t^{1/3}$ as in the case without elastic misfits \cite{96fra}.

When uniaxial external stress is included in the model,
wavy parallel stripes develop with an orientation depending on the direction
of applied stress \cite{95lab,97lab}.
Thus, the cubic symmetry in the domain morphology is broken,
in agreement with the experiments on rafting described in section \ref{raft}
Surprisingly, the mean domain size still increased in proportion to
$t^{1/3}$ in the model.

When the chemical interaction is taken to be a nearest neighbour
repulsion instead of an attraction \cite{98nie},
precipitates with atomic order (of antiferromagnetic type) are formed.
The qualitatively new feature compared to the previous cases is that the
precipitates can be 
ordered on two different atomic sublattices. Where two domains ordered on
different sublattices join, there is an anti-phase boundary
(APB). Such APBs have the property of being wetted by the disordered
matrix, thereby creating new boundaries between precipitates. The
presence of wetted APBs also influences the rate of 
coarsening. The resulting patterns, illustrated in Fig. 8, are extremely
similar to those 
found for nickel-base superalloys, as described in section \ref{morph} . 

\section{Conclusion}

We can now consider to what extent the theoretical models
we have been describing explain the various experimental phenomena
mentioned in section \ref{exp}. 

{\em The shapes of the precipitates}

The plate-like shape of the individual precipitates in alloys such as Al-Cu 
and Cu-Be, mentioned in section \ref{morph} is consistent
with the prediction of the sharp-interface model (section \ref{aniso-shape})
that when the interface energy is negligible and the material is 
elastically anisotropic
the equilibrium shapes of the precipitates are plates 
perpendicular to an elastically soft direction.
Two-dimensional multi-precipitate simulations 
with anisotropic elasticity,
using either the diffuse-interface model (see section \ref{DI-many} 
and Fig. 7(d,e)) or the
atomic lattice model \cite{96fra}, also give long thin precipitates
perpendicular to an elastically soft direction.

The dependence of the shape of precipitates on their size, 
(more precisely, on the parameter $R/R_0$ defined in section \ref{scope})
was investigated in the experiments with Ni-Al-Mo alloys illustrated in Fig. 1.
The individual precipitates looked round when 
this parameter was very small, but roughly square when it was large.
All three of the models we have described explain this dependence as an
effect of the elastic anisotropy;
it is seen particularly clearly in calculations using the 
sharp interface and diffuse interface models, where the behaviour
of a single precipitate as a function of its size (that is, of the
ratio of its elastic to its interfacial energy) has been 
considered (see sections \ref{aniso-shape}, \ref{diff-1} and \ref{DI-1}).

The splitting of large cube-like precipitates in some nickel-base
alloys (see Fig. 4) has also been investigated using various theoretical
approaches, but the results are inconsistent. 
Sharp-interface energy calculations assuming inclusions
of fixed shape do give this effect (sections \ref{aniso-shape}
and \ref{aniso-pair}, as do diffuse-interface calculations
of the equilibrium shape (sections \ref{DI-1}).
On the other hand the sharp-interface evolution calculations 
that have been done so far (section \ref{diff-1}) do not exhibit the effect. 

{\em The arrangement of the precipitates}

Experiments show that when the (anisotropic) elastic interaction 
is strong enough to affect the shapes of the precipitates, 
these precipitates tend to align along certain directions (see Fig 1(c)). 
A similar type of ordering is found in the sharp-interface model
(sections \ref{PC}, \ref{diff-2}, \ref{LSW}) and the diffuse-interface
model (section \ref{DI-many}).


{\em The rate of coarsening}

\cs While the $t^{1/3}$ growth law for coarsening in alloys without elastic
misfit 
is reasonably well understood \cite{61lif,61wag,89mul}, 
theory has been less successful in explaining the growth law
when elastic forces do act. 
The experiments illustrated in Fig. 3 indicate \cite{95par3} 
that if the elasticity is strongly anisotropic and (relatively) 
weakly heterogeneous, the coarsening follows a $t^{1/3}$ growth law, 
whereas if it is weakly anisotropic and strongly heterogeneous, 
the coarsening follows a $t^{1/3}$ law
only for a limited time, and then slows down or stops. 
The atomic lattice simulations described in Section \ref{ncfres}
also give $t^{1/3}$ growth for a purely anisotropic lattice, 
but theory does not explain
why the elastic forces apparently do not affect the growth exponent
at all in this case. For a heterogeneous lattice without anisotropy
the diffuse-interface simulations described in Section \ref{DI-many}
predicted that the growth would be slower than $t^{1/3}$ but not that
the growth exponent would change rather abruptly as shown in Fig. 3(b,c). 
Theory also predicts an interesting topological rearrangement
if in the initial configuration the precipitates are softer (less rigid)
than the matrix, but these experiments did not reveal any striking 
difference of behaviour between precipitates that were much softer 
than the matrix and ones that were harder. 

{\em Rafting}

The effect of an externally applied stress on the shapes and arrangement
of the precipitates, described in section \ref{raft} and 
illustrated in Fig. 1(g),  can be understood in principle on the
basis of isotropic (but inhomogeneous) elasticity using
the sharp interface model, as discussed in section \ref{elementary}. 
The effect has been studied in a more detailed way, 
both for  isotropic and anisotropic elasticity,
using all three of the theoretical approaches we have described
(sections \ref{iso-1}, \ref{DI-many}, \ref{cfres}, \ref{ncfres} and 
Fig. 8(c,d)). All these approaches reproduce, to a greater or less extent,
the qualitative experimental features of rafting, summarized in equation 
(\ref{raftsign}).

{\em Generalizations of the model}

Figs 2 and 8 illustrate two cases that go beyond the simple binary
alloy with either isotropic or homogeneous anisotropic elasticity
which underlies most of the theoretical
work described in this article. In Fig. 2 the precipitates have
a different crystal symmetry from the matrix, and
in Fig. 8 the precipitates can occur in more than one variant.
In both cases, the results illustrate the excellent qualitative
agreement with experiment that simulations based on 
a well-chosen theoretical model can give.

\section{Acknowledgements} 

We thank Istvan Gy\"{o}ngy and Armen Khachaturyan for helpful information,
Richard Weinkamer for his comments on an earlier draft of this paper,
and the Austrian Academy of Sciences for financial support.

\newpage

\section{List of symbols} 


\begin{tabular}{lll}

Symbol & first appearance       & meaning \\

$B(\bk)$ & (\ref{B})            &                                       \\
$\sB(\bk)$      & (\ref{Eel})   & `elastic potential'                   \\
$C_{11}, C_{12}, C_{44}$ & (\ref{cubic-w}) & elastic moduli of cubic crystal\\
$c$     &  (\ref{c-cons})       & concentration of A (solute) atoms     \\
$c_0^\alpha$    & sec. \ref{MD}   & 
        zero-temperature equilbrium value of $c$ in $\Omega_\alpha$     \\
$c_0$   & (\ref{Vegard})        & concentration at which stress-free strain
is 0 \\ 
$D$     & (\ref{diffneq})       & diffusivity                           \\
$E$     &  (\ref{iso-khachw})   & Young's modulus                       \\
$E(\bp)$& (\ref{FPF})           & elastic two-site interaction          \\
${e}_{ij}$      & (\ref{strain})& strain tensor                        \\   
${e}_{ij}^{0}$  & (\ref{w})     & stress-free strain tensor            \\   
$e_{ij}^{ext}$  & (\ref{sG})    & `externally applied strain'          \\
 
${e}_{ij}^\alpha$  & (\ref{ek}) & stress-free strain in phase $\alpha$ \\
$F$     &  (\ref{free-energy})  & free energy of system                 \\
$F^\Gamma$& (\ref{free-energy}) & free energy of interface              \\
$f$     &  (\ref{free-energy})  & thermodynamic free energy density     \\
${G}$   & (\ref{iso-w})         & shear modulus                         \\
$H_{ijmn}$      & (\ref{LRint}) &                                       \\
${Z}$   & (\ref{sG})            &                                       \\
$K$     &  (\ref{iso-w})        & bulk modulus                          \\
$K_*,K_\bullet$   &  (\ref{Kstar}), (\ref{Kbullet})        &
\\         
${\rm k}$       &               & Boltzmann's constant                  \\
$L$     &  (\ref{cfW})          & stiffness of a longitudinal spring    \\
$l$     &  (\ref{cfW})          & natureal length of a longitudinal spring \\
$M$     & (\ref{Cahnkinetic},\ref{Khacheq})      & mobility coeffient or
matrix  \\ 
$\bn$   &  various              & a unit vector                         \\
$p$     & (\ref{p})             & excess pressure inside inclusion      \\
$\bp$   & section \ref{Born}    & a lattice site                        \\
$q$     & (\ref{iso-SFS})       & isotropic stress-free strain          \\
$R$     &  section \ref{scope}  & radius of (spherical) precipitate     \\     
$R_0$   &  section \ref{scope}  & `crossover' value of $R$              \\
$r_{\bp,\bp'}$  &(\ref{cfW})    & linearized distance between atoms on
sites $\bp,\bp'$ \\ 
$ T$    &                       & temperature                           \\
$T$     & (\ref{transW})        & stiffness of a transverse spring      \\
$T_0$   & (\ref{meanfieldf})    & mean-field transition temperature     \\
$ T_i $ &  (\ref{deltaW})       & traction                              \\
$t$     &  (1)                  & time                                  \\
$t_{ij}$&  (\ref{stress})       & stress tensor                        \\
$t_{ij}^{ext}$&(\ref{Wext1})     & externally applied stress tensor      \\
$t_{axial}^{ext}$&(\ref{raftsign})& axial component of anisotropic part of
$t^{ext}$\\ 
\end{tabular}
\newpage
\begin{tabular}{lll}
$\bu$   &  (\ref{strain})       & displacement field                    \\
$\bu'$  &  (\ref{u'})           & periodic part of $\bu$                \\
$V$     &  (\ref{V})            & total effective two-site interaction  \\
$V_{chem}$&  (\ref{FPF})        & chemical interaction                  \\
$V_{el}$&  (\ref{Vel})          & two-point elastic interaction         \\
$v_n$   &  (\ref{Gammadot})     & normal velocity of interface          \\
$W$     &                       & elastic energy of the specimen        \\
$W^{ext}$& (\ref{free-energy})  & potential energy of loading mechanism \\
$W_{int}$& (\ref{EshelbyE})     & elastic interaction energy            \\
$w$     &  (\ref{free-energy})  & elastic energy density                \\
$\bar{w}_\alpha$ & (\ref{spherew})      & elastic energy per unit volume of
inclusion \\ 
$\bx$   &  (\ref{free-energy})  & reference position of material point  \\
$x_i$  & (\ref{strain})        & Cartesian components of $\bx~(i =
1,2,3)$\\                       
$\alpha,\beta$ &                & the A-rich, B-rich phases             \\
$\Gamma$&  (\ref{free-energy})  & the interface between phases $\alpha$ and
$\beta$\\     
$\gamma$& (\ref{cfW})            & $+1(-1)$ for site occupied by an $A(B)$
atom \\ 
$\Delta(\bz)$   & (\ref{V})     & $\chi$ times finite-difference Laplacian\\
$(\Delta{e})_{ij}$ & (\ref{iso-w})      & $e_{ij}-e_{ij}^{0}$           \\   
$\delta_{ij}$ & (\ref{iso-SFS}) & 1 if $i=j$, 0 if not                  \\
$\eta$  &  (\ref{Vegard})       & compositional expansion coefficient   \\
$\eta(\bx)$     & (\ref{AC})    & order-parameter field                 \\
$\theta(\bk)$ & (\ref{theta})   &                                       \\
$\kappa$& (\ref{deltaW})        & mean curvature of $\Gamma$            \\
 
$\Lambda$       & (\ref{AC})    & kinetic coefficient in Allen-Cahn eqn \\

$\lambda_{ijmn}$ & (\ref{w})    & elastic (stiffness) tensor            \\
$\mu$   & (\ref{sfmu})          & chemical potential                    \\
$\hat{\mu}$     & (\ref{hmu})   & diffusion potential                   \\
$\nu$   & (\ref{iso-khachw})    & Poisson's ratio                       \\
$\pi$   & (\ref{pi})            & grand canonical pressure              \\
$\sigma$ & (\ref{Fint})         & interfacial energy (surface tension)  \\
$\chi$  & (\ref{CahnF1})        & coefficient of gradient term in free
energy density \\ 
$\Psi_{ijmn}$& (\ref{Psi})      &                                       \\
$\Omega,\Omega_\alpha,\Omega_\beta $
  & (\ref{free-energy})    & region occupied by specimen, inclusions, matrix\\
$\partial\Omega$ & (\ref{Wext1})& surface of $\Omega$                   \\
$|\Omega|$      & (\ref{FTw})   & volume of $\Omega$                    \\
$O(\dots)$      & (\ref{picmu}) & quantity not much larger than $(\dots)$
\\        
$(\dots)^\alpha$&               & value of $(\dots)$ in $\Omega^\alpha$ \\
$\delta(\dots)/\delta(\dots)$   & (\ref{varderiv})      & variational (or
functional) derivative        \\ 
$(\dots)_i$     &               & $i$-component of the vector $(...)$   \\
$(\dots)_{,j}$  &(\ref{stress}) & $\partial (\dots) /\partial x_j$      \\
$[\dots]$& (\ref{ucont})        & $[\dots]^\alpha_\beta$                \\
$(\dots)^*$     &               & complex conjugate of $(\dots)$        \\
$\tilde{(\dots)}$& (\ref{FT})   & Fourier transform of $(\dots)$        \\
$\bar{(\dots)}$ & (\ref{FTw})   & space average of $(\dots)$            \\
\end{tabular}

\newpage

\section{Figure captions}

Fig. 1: (a) Transmission electron micrograph (TEM) of Ni-Al-Mo alloy with
Mo-composition chosen such as to make the lattice spacing in matrix and
precipitates (gamma-prime phase) equal. Treatment: 430h at 1048 K after
quench from
the single-phase region. The plane of observation corresponds to the
crystallographic plane (001). One observes round precipitates 
{\em i.e.} no elastic effects.\\
(b) Small-angle X-ray scattering (SAXS) data from the same specimen in the
same orientation. One observes an isotropic scattering pattern.\\
(c) TEM of the (001)-plane of a Ni-Al-Mo alloy with Mo-composition chosen
such as to make the lattice spacing in the precipitates (gamma-prime phase)
larger by 0.4
precipitates, aligned along the elastically soft directions, [010] and [100].\\
(d) SAXS from the specimen in (c). There is a flower-like pattern with
four-fold
symmetry, the elongations being in [010] and [100]-directions (indicating
flat interfaces and strong correlations in those directions).\\
(e) TEM of Ni-Al-Mo alloy with Mo-composition chosen such as to make the
lattice spacing in the precipitates (gamma-prime phase) smaller
(misfit -0.5
qualitatively similar to those in (c).\\
(f) SAXS from the specimen in (e).\\
(g) Same alloy and heat-treatment as in (e), but now with a external
compressive
load of 130MPa applied to it along the vertical [010]-direction.\\
(h) SAXS from the specimen in (g). Note that the horizontal streak has
disappeared, corresponding to the disappearence of vertically oriented
interfaces in (g).\\
The data in (a)-(d) are taken from \cite{95fah} and in (e)-(h) from
\cite{97lab}.

Fig. 2: (a) Chessboard-like microstructure observed by transmission
electron microscopy in Co$_{39.5}$-Pt$_{60.5}$ slowly cooled from 1023K
and aged 15 days at 873K. The scale bar represents 30 nm. The cubic
phase appears white while the tetragonal phase is black.
(b) \ds Computer simulation using a diffuse interface model 
(From \cite{98bou}).

Fig. 3: Time-dependence of the mean precipitate size  for (a) Ni-Al-Mo
alloys (from \cite{95fah}), (b) Ni-Cu-Si alloys (from \cite{87yos})
and (c) Ti-Mo alloys (from \cite{94lan}). The coefficients M(c) in (a)
and M(T) in (c) are used to scale data obtained for different alloy
compositions or different temperatures, respectively.

Fig. 4: Typical example of ordered precipitates in a nickel-base superally,
which split into several pieces during the coarsening process (from
\cite{94qu}).

Fig. 5. Geometrical objects used in section \ref{interfcond}

Fig. 6. A typical free-energy density function $f$.

Fig. 7: Typical snapshot pictures 
for isotropic but heterogeneous elasticity (a-c) and homogeneous
elasticity with cubic anisotropy (d-e) using equations of the type
(\ref{CahnF2}-\ref{eCH}). In (a-c), the stiffer phase is shown white
with volume fractions of 0.3, 0.5 and 0.7, respectively (from
\cite{91onu3}). Note that the softer phase always wraps stiffer
particles. In (d-e), the volume fraction of the white phase is 0.5 and
0.7, respectively (from \cite{90nis}).

Fig. 8: (a) and (c) are reproductions of Figs1(e) and 1(g), respectively.\\
(b) shows results from computer simulations of an Ising model with elastic
interactions on a simple square lattice and with repulsive interaction of
like
atoms on nearest neighbor sites (with interaction energy J) and attractive
interaction of like atoms on next nearest neighbor sites (energy J/2). Black
shows the disordered phase (containing mostly A-atoms) and white the ordered
phase (consisting of about half A and half B atoms). The overall
concentration of B-atoms was 0.35 and the run was performed at a temperature
of T=0.567J/k on a $128\times128$ lattice and stopped after $10^6$ MCS .\\
(d) The same model, temperature and annealing time as in (b), but with an
additional external load along the vertical direction (simulations
from \cite{98nie})

\newpage

\end{document}